\documentclass[prd,preprint,eqsecnum,nofootinbib,amsmath,amssymb,
               tightenlines,dvips]{revtex4}
\usepackage{graphicx}
\usepackage{bm}

\begin {document}


\def\ca{C_{\rm A}}
\def\cs{C_s}
\def\cf{C_{\rm F}}

\def\df{d_{\rm F}}
\def\da{d_{\rm A}}
\def\Nf{N_{\rm f}}
\def\Nc{N_{\rm c}}
\def\ta{t_{\rm A}}
\def\tf{t_{\rm F}}
\def\md{m_{\rm D}}
\def\half{\tfrac{1}{2}}
\def\Real{\operatorname{Re}}

\def\p{{\bm p}}
\def\q{{\bm q}}
\def\k{{\bm k}}
\def\x{{\bm x}}
\def\B{{\bm B}}
\def\b{{\bm b}}

\def\x{{\bm x}}
\def\v{{\bm v}}
\def\gammaE{\gamma_{\rm E}^{}}

\def\qhat{\hat{\bar q}}
\def\qhatA{\hat q_{\rm A}}
\def\qhatF{\hat q_{\rm F}}
\def\grad{{\bm\nabla}}
\def\tr{\operatorname{tr}}

\def\linf{l_\infty}
\def\Linf{L_\infty}
\def\lf{l_{\rm f}}

\def\Q{{\bm Q}}

\def\Re{\operatorname{Re}}


\title
    {
      High-energy gluon bremsstrahlung in a finite
      medium: harmonic oscillator versus
      single scattering approximation
    }

\author{Peter Arnold}
\affiliation
    {%
    Department of Physics,
    University of Virginia, Box 400714,
    Charlottesville, Virginia 22904, USA
    }%

\date {\today}

\begin {abstract}%
{%
  A particle produced in a hard collision can lose energy through
  bremsstrahlung.  It has long been of interest to calculate the
  effect on bremsstrahlung if the particle is produced inside
  a finite-size QCD medium such as a quark-gluon plasma.
  For the case of very high-energy particles traveling through
  the background of a weakly-coupled quark-gluon plasma,
  it is known
  how to reduce this problem to an equivalent problem
  in non-relativistic two-dimensional quantum mechanics.
  Analytic solutions, however, have always resorted to further
  approximations.  One is a harmonic oscillator approximation to
  the corresponding quantum mechanics problem, which is appropriate
  for sufficiently thick media.  Another is
  to formally treat the particle as having only a single
  significant scattering from the plasma (known as the $N{=}1$
  term of the opacity expansion), which is appropriate for sufficiently
  thin media.
  In a broad range of intermediate cases,
  these two very different approximations give
  surprisingly similar but slightly differing results
  if one works to leading logarithmic order in
  the particle energy, and there has been confusion about the range
  of validity of
  each approximation.
  In this paper, I sort out in detail the parametric
  range of validity
  of these two approximations at leading
  logarithmic order.
  For simplicity, I study the problem for
  small $\alpha_{\rm s}$ and large logarithms but
  $\alpha_{\rm s} \log \ll 1$.
}%
\end {abstract}

\maketitle
\thispagestyle {empty}


\section {Introduction and Results}
\label{sec:intro}

\subsection {Background}

There is a prototypical toy problem
often considered in
theoretical discussions of gluon bremsstrahlung
in a QCD medium such as a quark-gluon plasma:
Consider a high-energy quark or gluon that is produced by some
hard scattering event and then propagates through a length $L$ of
a uniform QCD medium before emerging into vacuum.
What is the effect of the medium on the probability for gluon
bremsstrahlung from this high-energy particle?
This is known as the brick problem.%
\footnote{
  There are other versions of this problem.  Sometimes people consider
  the case of a high-energy quark or gluon that propagates a relatively
  long distance through vacuum, then enters a uniform QCD medium of
  length $L$, passes through it, and exits the other side, approximately
  maintaining its direction throughout.  I instead consider the case
  where the particle is first created inside the medium.  Created could
  mean as one member of a particle/anti-particle pair well separated in
  angle (in which case one would separately compute the medium
  effect on bremsstrahlung from the other particle), or it could mean
  the final state of a large-angle deflection
  of a pre-existing particle (in which case one would
  separately compute the medium effect on initial-state radiation).
}
The problem is complicated by the Landau-Pomeranchuk-Migdal (LPM)
effect \cite{LP,Migdal}.
The quantum mechanical duration (formation time) of the
bremsstrahlung process grows with increasing energy and eventually
exceeds the mean free time between collisions.  As a result,
successive collisions of the high-energy particle with the plasma
cannot be treated as independent from each other for the purpose
of calculating the probability of bremsstrahlung.

There is a general formalism for treating this problem,%
\footnote{
   See \cite{BSZ,BDMPS1,BDMPS2,BDMPS3,Zakharov1,Zakharov2} for the
   original development.  See also \cite{timelpm1} for a summary in a
   language that generalizes naturally to the problem of non-fixed scatterers,
   and for a discussion of how the formalism is related to that
   developed in Refs.\ \cite{AMYsansra, AMYkinetic, AMYx}
   for the case of infinite media.
   For a nearly complete calculation of bremsstrahlung in the
   infinite medium case, to leading order in $\alpha$, see
   Ref.\ \cite{JeonMoore}.
}
but analytic solutions have required additional approximations.
Baier, Dokshitzer, Mueller, and Schiff (BDMS)%
\footnote{
  See also the earlier work with Peigne of Refs.\ \cite{BDMPS2,BDMPS3}.
}
\cite{BDMS}
investigated
the problem in the limit that
the energy was high enough, and
the medium thick enough, that the number of collisions $N$ within
the bremsstrahlung formation time was large --- so large that $\ln N$
could be treated as large.
In this limit, they made an
approximation, known as the harmonic oscillator (HO)
approximation, that reduced the general formalism to a
certain type of harmonic oscillator problem.
They solved for the medium effects on the
spectrum of gluon bremsstrahlung, to be reviewed below.
From the spectrum, they computed the
size $\Delta E$ of the medium effect on the average
energy loss
of a high-energy particle of energy $E$.%
\footnote{
  The single number given by the {\it average} energy loss leaves much to
  be desired as a description of the final-energy probability
  distribution because that distribution tends to have large,
  non-Gaussian tails.  See the discussion in Sec.\ 3 of Ref.\
  \cite{BDMSquench}
  or Ref.\ \cite{JeonMoore}.  However, here my purpose is just to use
  it as an example for the sake of theoretically comparing the roles
  of the HO and $N{=}1$ approximations.
}
The qualitative form of their result depends on the thickness $L$ of
the medium compared to the typical formation length $L_\infty$
for gluon bremsstrahlung in an infinite medium, which is parametrically
\begin {equation}
   L_\infty \sim \sqrt{\frac{E}{\hat q}} .
\label {eq:Linf}
\end {equation}
Here, $\hat q$ is the typical squared transverse momentum per unit
length transferred via elastic collisions to a high-energy particle as
it traverses the medium (more discussion later).
For thick media ($L \gg L_\infty$),
they found that $\Delta E$ grows linearly with $L$, as one would expect.
For thin media ($L \ll L_\infty$), they found \cite{BDMS}%
\footnote{
  Specifically, the first equality in (\ref{eq:dEHO}) is equivalent to
  Eq.\ (49) of Ref.\ \cite{BDMS}, which can be expressed in terms
  of $\hat q$ as $(\Delta E)/L = \tfrac14 \alpha \ca \hat q_s L$,
  where $\ca = \Nc$ for SU($\Nc$) gauge theory.
  Since $\hat q_s$ is proportional to $C_s$, one can rewrite this
  in the form $\tfrac14 \alpha \cs \qhatA L$, which will be more
  convenient for my later discussion.  The last equality in
  (\ref{eq:dEHO}) is given by the formula for $\hat q$, which I
  review later in (\ref{eq:qhat}).
}
\begin {equation}
   \Delta E_{\rm HO} \simeq
   \tfrac14 C_s \alpha \qhatA L^2
   \simeq
   \pi C_s \ca \alpha^3 {\cal N} L^2
   \ln\left( \frac{\qhatA L}{\md^2} \right) ,
\label {eq:dEHO}
\end {equation}
to leading order in inverse powers of the logarithm.
Here $s$ is the species (quark or gluon) of the high-energy particle,
and $C_R$ is the quadratic Casimir of a given color representation.
${\cal N}$ is the density $n$ of plasma particles weighted by
group factors as%
\footnote{
  Here $d_{\rm R}$ is the dimension of color representation $R$, $n_{\rm g}$
  is the total gluon density, and $n_{\rm q}$ and $n_{\rm \bar q}$ are
  the total quark and anti-quark densities, summed over flavor.
  One may equivalently write
  ${\cal N} = (\ta/\da) n_{\rm g} + (\tf/\df) (n_{\rm q} + n_{\rm \bar q})
  = 2 \ta n_+ + 4 \Nf \tf n_-$ where $t_R$ is the trace normalization
  defined in terms of color generators $T_R^a$ by
  $\tr(T_R^a T_R^b) = t_R \delta^{ab}$, and
  $n_\pm = \int (2\pi)^{-3}d^3p \> (e^{\beta p}\mp1)^{-1}$ is the
  number density of a single, massless, bosonic/fermionic degree of freedom.
}
\begin {equation}
   {\cal N} =
   \frac{1}{\da} \left[
     \ca n_{\rm g} + \cf (n_{\rm q} + n_{\rm \bar q})
   \right]
   =
     \tfrac38 \, n_{\rm g} + \tfrac16 (n_{\rm q} + n_{\rm \bar q})
   = \frac{6\,\zeta(3)}{\pi^2} \Bigl( 1 + \tfrac14 \Nf \Bigr) T^2 ,
\end {equation}
where $N_{\rm f}$ is the number of quark flavors.

In contrast, various other authors
have investigated the opposite approximation, starting
from early work by Wiedemann and Gyulassy \cite{WiedemannGyulassy}
and by Gyulassy, Levai, and
Vitev (GLV) \cite{GLV,GLV2a,GLV2b}.
Instead of treating the number $N$ of elastic collisions as large,
they expand
order by order in the number of collisions.  This is known as the
opacity expansion.  The leading term, corresponding to $N=1$, gives%
\footnote{
   Eq.\ (16) of Ref.\ \cite{GLV2a} gives
   $(C_R \alpha L^2 \mu^2/4\lambda_{\rm g}) \ln(E/\mu)$, where
   $\lambda_{\rm g}$ is the gluon mean free path and $\mu^{-1}$ is the
   color electric screening length.  The older literature on
   gluon bremsstrahlung often unnecessarily
   normalizes answers in terms of $\lambda_{\rm g}$,
   which is not well defined in leading-order perturbation
   theory because of a logarithmic infrared divergence from {\it magnetic}\/
   scattering.  But the result for bremsstrahlung does not depend
   on these details.
   Using the purely electric scattering models assumed in
   older calculations,
   $\mu^2/\lambda_{\rm g} = 4\pi \ca \alpha^2{\cal N}$.
   This substitution recovers independence from
   the details of electric vs.\ magnetic screening.
   The fact that the appropriate
   lower scale in the logarithm is of order $\md^2 L$ can be
   found in the work of Zakharov \cite{ZakharovResolution}
   and is nicely laid out in the
   presentation of Salgado and Wiedemann \cite{SalgadoWiedemann}.
   Alternatively, readers of
   GLV can see it by
   noting that Eq.\ (15) of Ref.\ \cite{GLV2a}
   (Eq.\ (130) of Ref.\ \cite{GLV2b})
   is only valid when $\gamma \ll 1$
   and so $x \gg L \md^2/E$, and so the
   infrared
   logarithmic divergence in the $x$ integration of that equation
   is cut off by this lower bound on $x$ and generates the
   logarithm in (\ref{eq:dEN1}) above.
}
\begin {equation}
   \Delta E_{N{=}1} \simeq
   \pi C_s \ca \alpha^3 {\cal N} L^2
   \ln\left( \frac{E}{\md^2 L} \right)
\label {eq:dEN1}
\end {equation}
in the high-energy limit,
to leading order in inverse powers of $\ln(E/\md^2 L)$.

The results (\ref{eq:dEHO}) and (\ref{eq:dEN1}) from opposite
assumptions about the relevant number of collisions are surprisingly
similar, differing only in the argument of the logarithm.
Two natural questions arise.  Which formula is correct for what
range of media thickness $L$?  Which description captures the
correct physics: Is there a single collision with the medium which
dominates the medium's contribution to bremsstrahlung energy loss,
or is $\Delta E$ dominated by processes where many scatterings
are important?

The qualitative difference between the HO and $N{=}1$ approximations
becomes more pronounced if one looks more generally at the gluon
bremsstrahlung spectrum instead of focusing on the single number
$\Delta E$.  For the brick problem, the HO approximation
gives the result \cite{BDMS}%
\footnote{
  For a relatively simple formula for more general situations of
  expanding, inhomogeneous media, see Ref.\ \cite{timelpm1}.
  See also the earlier work of Ref.\ \cite{BDMSc}.
  My sign convention in (\ref{eq:omega}) is that of Ref.\ \cite{timelpm1}.
}
\begin {equation}
   \omega \, \frac{d}{d\omega}(I-I_{\rm vac})_{\rm HO} =
   \frac{\alpha}{\pi} \, x \, P_{s{\to}{\rm g}}(x) \,
   \ln\left| \cos(\omega_0 L) \right|
\label {eq:IHO}
\end {equation}
with
\begin {equation}
   \omega_0^2 =
   - i \,
   \frac{[(1-x) \qhatA + x^2 \hat q_s]}{2 x (1-x) E} \,.
\label {eq:omega}
\end {equation}
Here, $I$ is the probability of gluon bremsstrahlung, with
$I_{\rm vac}$ the corresponding probability if the process which
created the high-energy particle had instead taken place in vacuum.
$P_{s\to g}(x)$ is the usual vacuum splitting function,%
\footnote{
   $P_{{\rm q}{\to}{\rm g}}(x) =
    \cf[1+(1-x)^2]/x$;
   $P_{{\rm g}{\to}{\rm g}}(x) =
    \ca[1+x^4+(1-x)^4]/x(1-x)$.
}
$\omega$ is the energy of the (high-energy) bremsstrahlung gluon,
and $x \equiv \omega/E$ is its momentum fraction.
$|\omega_0|^{-1}$ is of order the formation length
$l_\infty(\omega)$ for a bremsstrahlung gluon of frequency $\omega$
in an infinite medium.  The previous formula (\ref{eq:dEHO})
for $\Delta E$ is just the $\omega$ integral of (\ref{eq:IHO})
in the limit $L \ll L_\infty$.  In that limit, the $\omega$ integral
is dominated by small $x$ such that $l_\infty(\omega) \sim L$.
But now fix $\omega$ and consider thinner and thinner media
such that $L$ is small compared to the formation length
$l_\infty(\omega)$.  The small $L$ limit of (\ref{eq:IHO})
is
\begin {equation}
   \omega \, \frac{d}{d\omega}(I-I_{\rm vac})_{\rm HO} \simeq
   \frac{\alpha}{\pi} \, x \, P_{s{\to}{\rm g}}(x) \,
   \frac{|\omega_0 L|^4}{12}
\end {equation}
Focusing on the case $x\ll 1$, for simplicity, gives
\begin {equation}
   \omega \, \frac{d}{d\omega}(I-I_{\rm vac})_{\rm HO} \simeq
   \frac{C_s \alpha \qhatA^2}{24\pi \omega^2} \, L^4
   \simeq
   \frac{2\pi C_s \ca^2 \alpha^5 {\cal N}^2}{3 \omega^2} \, L^4
   \ln^2\left( \frac{\qhatA L}{\md^2} \right) ,
\label {eq:spectrumHO}
\end {equation}
where the last equality relies on the formula for $\hat q$, to
be reviewed momentarily.
In contrast, the $N{=}1$ term of the opacity expansion gives
\begin {equation}
   \omega \, \frac{d}{d\omega}(I-I_{\rm vac})_{N{=}1} \simeq
   \frac{\pi C_s \ca \alpha^3 {\cal N}}{\omega} \, L^2 .
\label{eq:spectrumN1}
\end {equation}
Because it is proportional to $L^2$ rather than $L^4$, the
$N{=}1$ result clearly dominates over the HO result for
small $L$.  The HO result seems completely at odds with
the $N{=}1$ result for small $L$.
When is the HO result correct?

All of these issues were raised some years ago by Zakharov
\cite{ZakharovResolution}.
He concluded that the HO analysis should not be trusted
in cases when the medium thickness $L$ is less than or order the
relevant formation length $L_\infty$ or $l_\infty(\omega)$.
In this paper, I return to this problem and show that the HO
approximation is valid over a wider range of $L$, and I elucidate
in more detail the interplay between contributions to bremsstrahlung
(i) arising from large numbers of scatterings and (ii) dominated by a
single scattering.

I will assume that the particle energy $E$ is so large
that $\ln(E/T)$ can be treated as a large number, where $T$ is the
plasma temperature.  Because I want to pursue a qualitative understanding
of how the HO and $N{=}1$ results fit together, I will usually just
focus on the parametric form of formulas.
Though I will treat
logarithms as large, I will formally assume that $\alpha$ is so small
that $\alpha \ln(E/T)$ is small.  So, for instance, I will ignore
running of the coupling and treat $\alpha$ as fixed.%
\footnote{
  For some discussion of running coupling in the bremsstrahlung problem,
  see for example Sec.\ VI of Ref.\ \cite{ArnoldDogan}, which combines earlier
  observations of Refs.\ \cite{BDMPS3} and \cite{Peshier}.
}
The purpose of these various limits is to provide a clean,
theoretical situation for conceptually disentangling the HO and $N{=}1$
approximations.
For
experimentally achievable quark-gluon plasmas, of course,
logarithms are not huge and $\alpha$ is not tiny.%
\footnote{
   There is some theoretical information concerning the efficacy of
   expansions in $1/\ln(E/T)$ in the context of
   infinite-medium bremsstrahlung calculations for weakly-coupled plasmas.
   In Ref.\ \cite{ArnoldDogan}, it was found that the error of making
   a next-to-leading logarithm approximation is
   $\lesssim 20\%$ when $E \gtrsim 10\,T$.
} 


\subsection {Transverse momentum diffusion}

To describe my results, it is useful to first
characterize the total transverse momentum $Q_\perp$ that a high-energy
particle picks up,
due to screened Coulomb-like interactions,
as it crosses length $L$ of a QCD medium.
In a perturbative quark-gluon plasma, the
differential elastic scattering rate for a high-energy particle
to pick up transverse momentum $q_\perp$ is
\begin {equation}
   \frac{d\Gamma_{\rm el}}{d^2q_\perp}
   \simeq \frac{4C_R \alpha^2{\cal N}}{q_\perp^4} \,,
\label {eq:dGammael}
\end {equation}
for $q_\perp \gg T$.  The behavior is similar, with a slightly
different coefficient, for smaller $q_\perp$ down to $\md$, where
Debye screening kicks in.  Over multiple collisions, the
net transverse momentum transfer will random walk,
and its average will be
\begin {equation}
   (Q_\perp^2)_{\rm avg}
   =
   L \int d^2q_\perp \frac{d\Gamma_{\rm el}}{d^2q_\perp} q_\perp^2 .
\label {eq:Qperp}
\end {equation}
Using (\ref{eq:dGammael}), this integral has a well-known logarithmic
divergence which is cut off in the infrared by Debye screening.
The UV end is cut off by the kinematic limit
$q_\perp^{\rm max} \sim \sqrt{E T}$, where the plasma temperature $T$ gives
the typical energy of a plasma particle.
However, the average $Q_\perp^2$ will not be an interesting quantity for
our purposes.  The probability that there is at least one collision with
individual momentum transfer of order $q_\perp$ over the distance
$L$ is of order
\begin {equation}
   \min\left(q_\perp^2 \frac{d\Gamma_{\rm el}}{d^2q_\perp}\,L \, , 1 \right)
   \sim
   \min\left( \frac{C_R \alpha^2 {\cal N} L}{q_\perp^2}\, , 1 \right)
\end {equation}
So it is unlikely to have any individual collisions with
$q_\perp^2 \gg C_R \alpha^2 {\cal N} L$.  If we are interested in
the {\it typical} (i.e.\ median) $Q_\perp^2$ instead of the
average $Q_\perp^2$, we should use this value of $q_\perp^2$
as an upper cut-off on the integration in (\ref{eq:Qperp}).%
\footnote{
  This important distinction of the typical or characteristic 
  $Q_\perp^2$ as opposed to the average was made previously in
  Sec.\ 3.1 of Ref.\ \cite{BDMPS3}.
}
The result is then
\begin {subequations}
\label{eq:Qtyp}
\begin {equation}
   (Q_\perp^2)_{\rm typ}
   =
   \hat q_R L
\end {equation}
with%
\footnote{
   In the context of infinite media, see
   Refs.\ \cite{ArnoldXiao,Simon} for the weak-coupling
   evaluation of $\hat q$ beyond leading
   log order and its application to bremsstrahlung calculations.
   See also the related, earlier work of Ref.\ \cite{BMT}.
}
\begin {equation}
   \hat q_R \simeq
   4\pi C_R \alpha^2 {\cal N}
   \ln\left( \frac{C_R \alpha^2 {\cal N} L}{\md^2} \right)
\label {eq:qhat}
\end {equation}
\end {subequations}
for a particle with color representation $R$.
Throughout this paper, I will use $\hat q$ to denote the
{\it typical} (rather than average) squared transverse momentum
acquired per unit length, as given by (\ref{eq:qhat}).

A simple way that one can get to an equivalent result is to
self-consistently use $Q_\perp$ itself to cut off the UV
logarithmic divergence in (\ref{eq:Qperp}), so that
\begin {equation}
   (Q_\perp^2)_{\rm typ}
   \simeq
   4\pi C_R \alpha^2 {\cal N} L
   \ln\left(\frac{(Q_\perp^2)_{\rm typ}}{\md^2} \right) .
\end {equation}
This gives the same result as (\ref{eq:Qtyp}) up to corrections that
are subleading in inverse powers of the logarithm.%
\footnote{
   Throughout this paper, I will treat large logarithms as
   parametrically large, but I will treat logarithms of
   logarithms, such as $\ln\ln(Q_\perp^2/\md^2)$, as being of order 1.
}
We can also write it in the form
\begin {equation}
   (Q_\perp^2)_{\rm typ}
   \simeq
   4\pi C_R \alpha^2 {\cal N} L
   \ln\left( \frac{\hat q_R L}{\md^2} \right) .
\end {equation}

\begin {figure}
\begin {center}
  \includegraphics[scale=0.6]{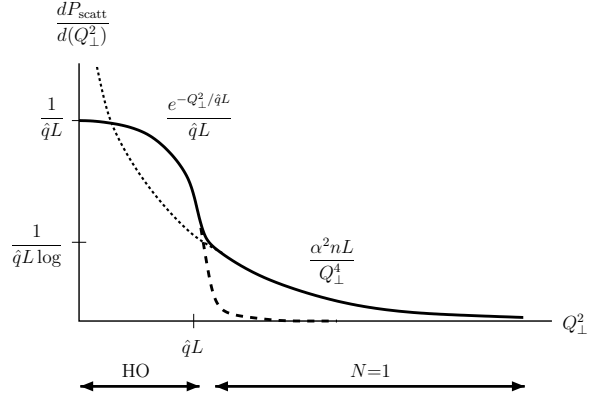}
  \caption{
     \label{fig:Pscatt}
     Probability distribution
     for a high-energy particle to pick up total transverse momentum
     $Q_\perp$ via
     $2\to2$ scattering while traveling through a medium
     of thickness $\sim L$.
     All formulas show only parametric dependence.
     The dashed line shows the behavior of the
     HO approximation at
     large $Q_\perp$, and the dotted line shows the
     behavior of the single-scattering
     ($N{=}1$) approximation at low $Q_\perp$.
     The kinematic upper limit $Q_\perp^2 \sim ET$ of the
     high-$Q_\perp$ tail is not shown.
  }
\end {center}
\end {figure}

A qualitative sketch of the
probability distribution of total $Q_\perp^2$ is shown by
the solid line in
Fig.\ \ref{fig:Pscatt}.
Typical events show a Gaussian peak characteristic of diffusion
in $\Q_\perp$ space, whose width is given by (\ref{eq:Qtyp}).
However, there is also a large $Q_\perp$ tail of rare events,
where one of the collisions with the medium has
$q_\perp \gg (Q_\perp)_{\rm typ}$.  The probability distribution
for these events is simply given by $d\Gamma_{\rm el}/dq_\perp^2$
times $L$.
The formulas shown in the figure are all parametric
and do not show multiplicative factors of $O(1)$.  
They also do not show group factors such as $C_{\rm R}$.
The normalization
$1/\hat q L$ of the height of the diffusion peak can be determined
from the requirement that the total probability is 1.  By comparing
the single scattering formula of the high-$Q_\perp$ tail and
the Gaussian formula for a diffusion peak, one can parametrically
estimate that the transition between the two occurs when
the probability distribution is down from the peak value by a factor
of order
\begin {equation}
   \log \equiv \ln\left( \frac{\hat q L}{\md^2} \right)
        \sim \ln\left( \frac{C_R \alpha^2 {\cal N}L}{\md^2} \right) .
\label {eq:log}
\end {equation}
I will use the short-hand notation ``$\log$'' defined above
to denote this particular logarithm in figures.
Further review of the important aspects of Fig.\ \ref{fig:Pscatt}
is given in appendix \ref{app:Pscatt} for readers desiring
more detailed explanation.

\begin {figure}
\begin {center}
  \includegraphics[scale=0.4]{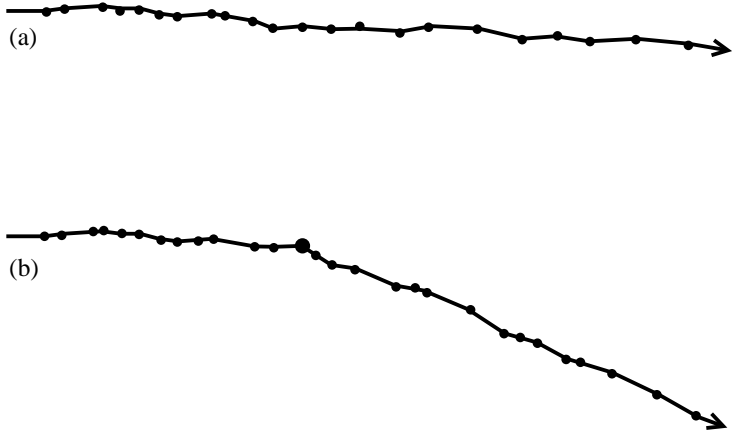}
  \caption{
     \label{fig:scatt}
     A depiction of (a) typical multiple scattering (the HO approximation)
     vs.\ (b) the rarer case (related to the $N{=}1$ approximation)
     where the total deflection is dominated by a
     single scattering with unusually large momentum transfer.
     The scattering angles are all exaggerated in this figure for
     the sake of visibility.  The net angular
     deflection in both cases is parametrically $\ll 1$, and the
     small-angle scatterings are meant to be significantly smaller
     than the single larger (but still small) angle scattering in
     (b).
  }
\end {center}
\end {figure}

The HO approximation corresponds to ignoring the large-$Q_\perp$
tail of this distribution and approximating the probability
distribution as a standard diffusion Gaussian peak,
\begin {equation}
   \frac{dP_{\rm scatt}}{d(Q_\perp^2)} \biggr|_{\rm HO}
   =
   \frac{1}{(Q_\perp^2)_{\rm typ}}
   \exp\left[
      - \frac{Q_\perp^2}{(Q_\perp^2)_{\rm typ}}
   \right] ,
\end {equation}
depicted
qualitatively by the dashed line in the figure.  The $N{=}1$
approximation, in contrast, involves using the single scattering
formula $\propto 1/Q_\perp^4$ for all momenta, all the way down
to the Debye mass.  This is depicted by the
dotted line in the figure, but the low-momentum
cut-off at $Q_\perp^2 \sim \md^2$
is not shown.  The double arrows beneath the plot indicate over which
regions these two approximations are good approximations to the
actual distribution.
A cartoon of a {\it typical} scattering is shown in
Fig.\ \ref{fig:scatt}a.  In contrast, a corresponding cartoon of one of
the rare high-$Q_\perp$ events is shown in Fig.\ \ref{fig:scatt}b.
Note that there are still many scatterings in this case, but a
single one of those scatterings dominates $Q_\perp$.
I will assume throughout this paper that the medium is thick enough
that the high-energy particle undergoes many soft collisions on
its way through, corresponding to Fig.\ \ref{fig:scatt}a or
Fig.\ \ref{fig:scatt}b.  Parametrically, this assumption is that
$L \gg (C_R \alpha T)^{-1}$.

The probability of having more than one scattering with
$Q_\perp \gg (Q_\perp)_{\rm typ}$
is a parametrically small correction to Fig.\ \ref{fig:Pscatt} and
so need not be considered.

Throughout this paper, I will assume that energies are high enough that
scattering and bremsstrahlung can be treated as nearly collinear.
In particular, I will restrict consideration to the case
$\omega \gg Q_\perp$.


\subsection {Results}

In this paper, I
will show that the leading log result for
$\Delta E$ gets two different types of contributions
when $L \ll L_\infty$:
\begin {equation}
   \Delta E \simeq
   \pi C_s \ca \alpha^3 {\cal N} L^2
   \left[ 
     \ln\left( \frac{\qhatA L}{\md^2} \right)
     + \ln\left( \frac{E}{\qhatA L^2} \right)
   \right] .
\label {eq:dE}
\end {equation}
The first logarithm is just the HO approximation of (\ref{eq:dEHO}),
corresponding to bremsstrahlung involving typical scattering from
the medium.
The second logarithm is due to events involving the rarer
scatterings corresponding to the large $Q_\perp$ tail of
Fig.\ \ref{fig:Pscatt}.  Amusingly, the sum of these
two logarithms simply gives the same mathematical formula as the
full $N{=}1$ result of (\ref{eq:dEN1}).  However, depending
on $L$, (\ref{eq:dE}) can be dominated by the HO contribution.
The formula (\ref{eq:dE}) is qualitatively similar to a result
by Zakharov \cite{ZakharovResolution},%
\footnote{
  Specifically, see Eq.\ (23) of Ref.\ \cite{ZakharovResolution}.
  This result is only qualitatively similar to my (\ref{eq:dE})
  because of the upper limit $\omega_{\rm cr}$ on Zakharov's
  HO term, which in my notation he takes of order
  $\omega_{\rm cr} \sim \hat q L^2$.  The parametric treatment
  of this cut-off means that his HO contribution is only parametrically
  of order the HO result at leading-log order (which is all he asserts
  in his text).  In my (\ref{eq:dE}),
  the first term is exactly the HO result at leading-logarithm order.
}
but Zakharov's conclusion about the domain of applicability
of the HO approximation was slightly different.

If one thinks of taking the high-energy limit $E \to \infty$ with
fixed $L$, then the $N{=}1$ term in (\ref{eq:dE}) obviously
dominates.  But now consider fixing a large value of $E$ and
varying $L$.  It's useful to rewrite (\ref{eq:dE}) parametrically
in terms of the typical formation length $L_\infty$ of
(\ref{eq:Linf}):
\begin {equation}
   \Delta E \sim
   C_s \ca \alpha^3 {\cal N} L^2
   \left[ 
     \ln\left( \frac{\sqrt{\qhatA E}}{\md^2} \frac{L}{L_\infty} \right)
     + \ln\left( \frac{L_\infty^2}{L^2} \right)
   \right] .
\end {equation}
For $L$ kept equal to any fixed fraction of $L_\infty$,
now the HO contribution dominates as $E \to \infty$.
The two logarithms are equal when $L$ is of order
\begin {equation}
   L_* \sim \left( \frac{\md^4}{\qhatA E} \right)^{1/6} L_\infty .
\label {eq:Lstar}
\end {equation}
Note that $L_*$ is small compared to the typical formation length
$L_\infty$ in the high energy limit.
For $L$ extremely large compared to $L_*$, the result for $\Delta E$ will
be dominated by the HO approximation, corresponding to scatterings like
Fig.\ \ref{fig:scatt}a.  In this limit, the formulas (\ref{eq:dEHO})
and (\ref{eq:dEN1}) in fact generate equivalent answers (that is,
their difference is small compared to the result).
For $L$ extremely small compared to $L_*$, $\Delta E$ will be
will be dominated by bremsstrahlung involving the large-$Q_\perp$
tail of Fig.\ \ref{fig:Pscatt} and so by scattering like
Fig.\ \ref{fig:scatt}b.  As I shall discuss, in this limit
the physics of bremsstrahlung is effectively single scattering
physics, even though there are multiple additional soft scatterings depicted
in Fig.\ \ref{fig:scatt}b.

\begin {figure}
\begin {center}
  \includegraphics[scale=0.6]{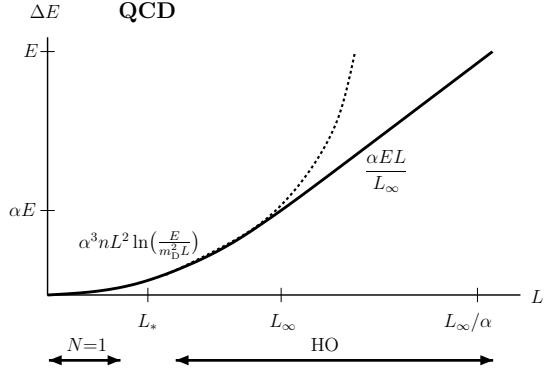}
  \caption{
     \label{fig:qcdEL}
     Total medium modification to QCD energy loss $\Delta E$,
     shown vs.\ medium length $L$.}
\end {center}
\end {figure}

The parametric results for $\Delta E$ are depicted qualitatively in
Fig.\ \ref{fig:qcdEL}, where the dotted line denotes the full
$N{=}1$ formula (\ref{eq:dEN1}), and the double arrows below the
graph again indicate whether the physics of the underlying elastic
scattering is dominated by the HO or $N{=}1$ type events of
Fig.\ \ref{fig:scatt}.
One consequence is that the HO approximation remains valid
when $L \sim L_\infty$.

I should clarify that the scales of the axis in my figures are
elastic and should not be interpreted as linear, though they do
start at zero in the bottom-left corner.

\begin {figure}
\begin {center}
  \includegraphics[scale=0.6]{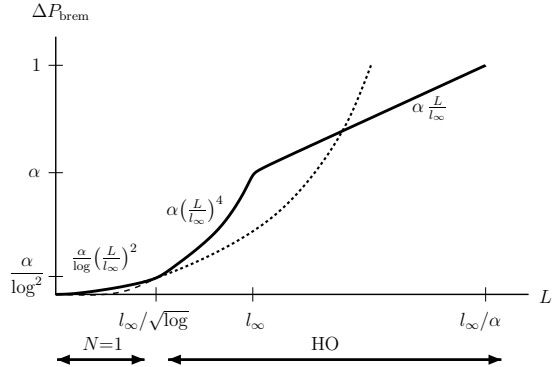}
  \caption{
     \label{fig:qcdPbremL}
     Medium modification to probability
     for emitting a high-energy bremsstrahlung gluon of
     frequency $\sim\omega$: total probability vs.\
     medium length $L$.}
\end {center}
\end {figure}

I will also preview my results concerning whether the bremsstrahlung
gluon spectrum decreases as $L^4$ (the HO prediction) or
$L^2$ (the $N=1$ prediction)
for small $L$ and fixed gluon frequency $\omega$.
Consider
\begin {equation}
   \Delta P_{\rm brem} \equiv
   \omega \, \frac{d}{d\omega} (I-I_{\rm vac}) ,
\label {eq:DPbremdef}
\end {equation}
which parametrically is
the medium
effect on the probability of bremsstrahlung
production of a gluon with frequency of order $\omega$.
Fig.\ \ref{fig:qcdPbremL} gives a qualitative sketch of
my result for $\Delta P_{\rm brem}$
versus the
medium length $L$.  In this figure, $l_\infty$ is short-hand for
the formation length $l_\infty(\omega)$ for gluons of that frequency.
For simplicity, I restrict attention to the case where
$1{-}x$ is not small, in
which case
\begin {equation}
   l_\infty(\omega) \sim \sqrt{\frac{\omega}{\qhatA}} \,.
\label {eq:linfQCD}
\end {equation}
Follow the $L$ axis from right to left.
As $L$ drops below $l_\infty$, the curve follows the
$L^4$ behavior of the HO approximation (\ref{eq:spectrumHO}).
At $L \sim l_\infty/\sqrt{\log}$, this corresponds
to a drop of $1/\log^2$ in probability from $L \sim \linf$.
As $L$ drops below $l_\infty/\sqrt{\log}$, the $L^2$ behavior of
the $N{=}1$ approximation takes over.

It's useful to note that one could generate the
entire $L \lesssim l_\infty$ behavior of Fig.\ \ref{fig:qcdPbremL} from
the HO and $N{=}1$ results (\ref{eq:spectrumHO}) and
(\ref{eq:spectrumN1}) if one made the assumption that the larger result
is the correct one.

Once again, one consequence is that the HO
approximation remains valid when $L$ is of order the relevant formation
time, treating logarithms as large.
However, in this particular case (unlike $\Delta E$),
the length scale at which the
$N{=}1$ result takes over is smaller by only a square root of a
logarithm.

In the remainder of this paper, I derive and explain these results.
In the next section, I will start with bremsstrahlung in QED rather than
QCD plasmas.  In particular, in section \ref{sec:basic}, I outline
how the typical and rare scattering events of Figs.\ \ref{fig:scatt}a
and b can both give potentially important contributions to the total
bremsstrahlung rate.  It's then a matter of discovering which is
the most important.  Next I briefly review the scales associated with
the LPM effect and then follow with detailed parametric estimates
of the relative importance of the different cases.  I move on to
QCD in section \ref{sec:QCD}, which requires relatively minor modifications
to the QED analysis, although the final results for energy loss as
a function of medium thickness are qualitatively quite different.
The conceptually most important result in this development will be
Fig.\ \ref{fig:qcdPbremQ}
(or Fig.\ \ref{fig:qedPbremQ} in the case of QED),
which shows the relative importance of
typical and rare scattering events to bremsstrahlung.
Finally, in section \ref{sec:Zakharov}, I reconcile the
results of this paper with earlier analysis by Zakharov
\cite{ZakharovResolution}.
Some more detailed arguments concerning some of the qualitative
points in this paper for QED and QCD bremsstrahlung are left to
Appendix \ref{app:qed} and \ref{app:qcd}, respectively.

Throughout the main text, I will focus on gluon bremsstrahlung
$s \to {\rm g}s$ with $x < 1/2$.
For the case of ${\rm g} \to {\rm g}{\rm g}$, that's everything
because of the identity of the final state particles.
For ${\rm q} \to {\rm g}{\rm q}$, however, there is an additional
contribution to energy loss from $1/2<x<1$.  The significance of this
contribution depends on whether one considers $\Delta E$ to be
the energy lost by the quark or the energy lost by the leading parton.
The thin-media formulas (\ref{eq:dEHO}) and (\ref{eq:dEN1}) use the
latter definition.  In Appendix \ref{app:xto1}, I spell out the details
and explain the simple way in which
Fig.\ \ref{fig:qcdPbremL} changes in the limit $x \to 1$.

Some of the parametric formulas I will derive in this paper were
derived earlier by BDMS \cite{BDMSquench}.  However, they did not keep careful
track of logarithms when comparing HO and $N{=}1$ contributions, which
is important for a discussion of disentangling which is the most
important at the order of leading logarithms.
The situation is also one of many topics discussed in a recent
mini-review by Peign\'e and Smilga \cite{PeigneSmilga}, who discuss energy
loss in both QED and QCD plasmas but do not keep track of logarithms.


\section{Bremsstrahlung in QED}

In this section, I focus on QED plasmas, deferring the
treatment of QCD until section \ref{sec:QCD}.

\subsection {Basic Picture}
\label {sec:basic}

\begin {figure}
\begin {center}
  \includegraphics[scale=0.3]{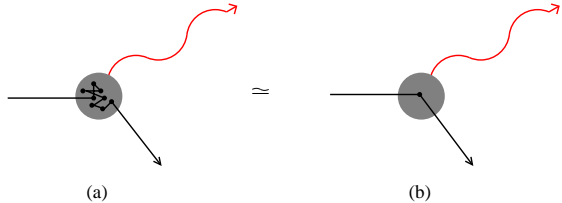}
  \caption{
     \label{fig:NR}
     A non-relativistic example of the probability for soft bremsstrahlung
     radiation being unable to resolve details of charge particle
     tracks that are smaller than the photon wavelength.
  }
\end {center}
\end {figure}

\begin {figure}
\begin {center}
  \includegraphics[scale=0.3]{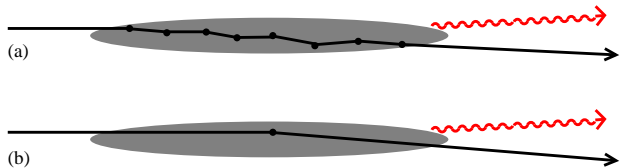}
  \caption{
     \label{fig:NRboost}
     The same processes as
     Fig.\ \ref{fig:NR}, now viewed from a highly boosted reference frame.
     The deflection angles are very small and the bremsstrahlung nearly
     collinear, but I have exaggerated the angles for the purpose of
     drawing the picture.
  }
\end {center}
\end {figure}

It's useful to first think about the case of
soft bremsstrahlung, for which the charged particle can be approximated
as classical.  Imagine that
Fig.\ \ref{fig:scatt} corresponds to possible particle tracks,
and we want to estimate the bremsstrahlung probability.
As a reminder of the origin of the LPM effect, first consider the
case of a non-relativistic particle and recall that light cannot
resolve features smaller than its wavelength.  Thus, bremsstrahlung
from the track in Fig.\ \ref{fig:NR}a will look the same as
that from Fig.\ \ref{fig:NR}b, provided the wavelength is large
compared to the distance scale over which the two trajectories
behave differently.  If one now Lorentz boosts this situation
to extremely high energy, then the size of the region the photon
cannot resolve will grow by a Lorentz factor and is now called the
photon formation length, while the photon
wavelength in that direction shrinks by a Lorentz factor.
The photon therefore cannot resolve the difference between the
situations of Fig.\ \ref{fig:NRboost}a and \ref{fig:NRboost}b.

In the high energy case, photon bremsstrahlung will be nearly collinear
with the charged particle, which can be understood as a result of the
boost.
In the ultra-relativistic limit, closer collinearity means larger
formation lengths.
Another useful mnemonic to keep in mind is that a photon emitted
at angle $\theta$ from an ultra-relativistic particle is not very
sensitive to particle deflections small compared to $\theta$.
So the photon emission angle will be less than or order the net
deflection angle of the charged particle within the formation length.

\begin {figure}
\begin {center}
  \includegraphics[scale=0.3]{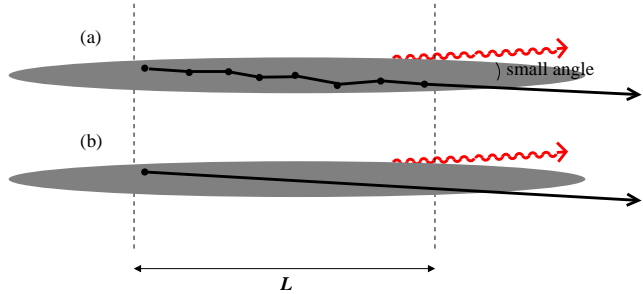}
  \caption{
     \label{fig:NRboost1}
     The equivalence of bremsstrahlung (a) with and (b) without medium
     interactions for the case where the length $L$ of medium traversed
     is small compared to the formation length.
  }
\end {center}
\end {figure}

Fig.\ \ref{fig:NRboost1} is somewhat similar to Fig.\ \ref{fig:NRboost}
but shows the case of propagation through a medium whose size
is small compared to the formation length.
The start of the particle trajectory corresponds to whatever hard
process (not shown) originally launched the high-energy particle in its
approximate direction of motion.  Bremsstrahlung photons cannot
resolve the difference between Fig.\ \ref{fig:NRboost1}a and
Fig.\ \ref{fig:NRboost1}b, and so the medium does not have
a significant effect on the probability of bremsstrahlung.

\begin {figure}
\begin {center}
  \includegraphics[scale=0.3]{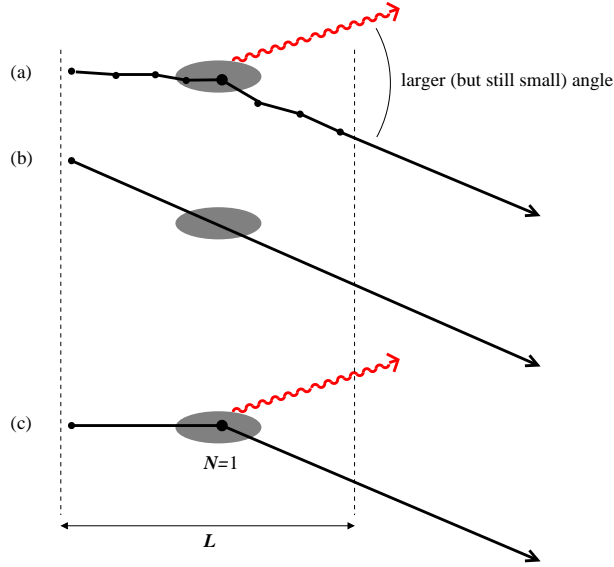}
  \caption{
     \label{fig:NRrare}
     Like Fig.\ \ref{fig:NRboost1}, but now (a) there is a rare scattering
     with larger than typical deflection angle, which affects
     bremsstrahlung at larger angles, which corresponds to
     smaller formation length (shaded ovals).  Case (b)
     is no longer approximately (a), but case (c) is.
  }
\end {center}
\end {figure}

Now consider the case
of rarer collisions that involve a single larger-than-typical
elastic scattering, as in Fig.\ \ref{fig:NRrare}a.
This scattering can then affect
bremsstrahlung radiation at larger than usual angles,
corresponding to photons with smaller formation
times.  If the angle is large enough, so that the formation length
in that particular case becomes smaller than the $O(L)$ distance
between the rare collision and the start of the particle's trajectory,
then the photon can resolve the difference between
Fig.\ \ref{fig:NRrare}a and \ref{fig:NRrare}b.
The rare collision is therefore a second chance for bremsstrahlung,
independent of the original event that produced the high-energy
particle and without any LPM suppression.
As far as this photon is concerned, the process is similar to
the $N{=}1$ process shown in Fig.\ \ref{fig:NRrare}c.

\begin {figure}
\begin {center}
  \includegraphics[scale=0.3]{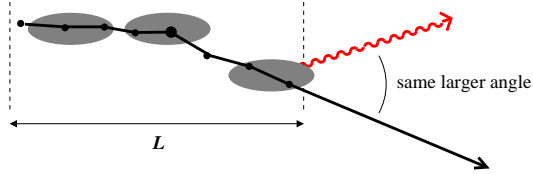}
  \caption{
     \label{fig:NRextra}
     As Fig.\ \ref{fig:NRrare} but considering the possibility of
     similar photon production from other, typical collisions
     along the trajectory.  This possibility is suppressed.
  }
\end {center}
\end {figure}

Because of the relatively small formation length, some readers may
wonder if the additional {\it typical}-angle scattering events in
Fig.\ \ref{fig:NRrare}a can provide additional, distinct opportunities
for bremsstrahlung, as depicted in Fig.\ \ref{fig:NRextra}, and
so ruin the equivalence of Figs.\ \ref{fig:NRrare}a
and \ref{fig:NRrare}c.  This does not happen because the
short-formation-length photons we have considered in the rare
scattering case of Fig.\ \ref{fig:NRrare} are emitted at angles large
compared to the {\it typical}\/ net scattering angle during one such
formation length.  The scatterings shown in the two additional ovals
of Fig.\ \ref{fig:NRextra}, for example, do not produce significant
bremsstrahlung radiation at such large angles.

The upshot of this discussion
is that scattering with larger than usual angles is
rarer but, when it does happen, the probability for an
associated bremsstrahlung photon is higher because there is less
LPM suppression.  Which type of process dominates the medium
effect on bremsstrahlung depends on which of these opposing
effects on probability is the most important.


\subsection {Review of LPM Effect}

\begin {figure}
\begin {center}
  \includegraphics[scale=0.3]{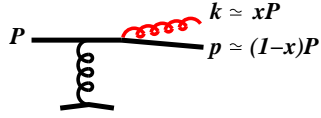}
  \caption{
     \label{fig:brem}
     One of the diagrams contributing to simple bremsstrahlung from
     a single scattering from the plasma.  (The other important diagram
     is the one where the bremsstrahlung comes from the initial
     high-energy particle line.)
 }
\end {center}
\end {figure}

For QED, one of the usual approaches to qualitative estimates of
the formation length is the following:%
\footnote{
  For a nice, very brief review, see, for example, the introduction
  of Ref.\ \cite{GyulassyWang}.
}
two space-time points $X_1$
and $X_2$ on the charged particle's trajectory lie within one
formation length if the relative phase $K_\mu(X_1-X_2)^\mu$ for
photon emission from those two points is $\ll 1$.  If the
particle is moving nearly linearly at close to the speed of light,
this condition becomes $\omega \, |\x_1{-}\x_2| (1 - \cos\theta) \ll 1$,
where $\theta$ is the angle between the photon and the charged particle.
Changing $\ll$ to $\sim$ then qualitatively defines the formation
length $\lf$, which for small $\theta$ gives
$\omega \lf \theta^2 \sim 1$ and so
\begin {equation}
   \lf \sim \frac{1}{\omega \theta^2} .
\label {eq:lpm1}
\end {equation}
A more general way
to the same result is to consider how off-shell in energy the
intermediate particle line is in a simple bremsstrahlung
diagram like Fig.\ \ref{fig:brem}, which is
\begin {equation}
   \delta E \equiv E_s(\p) + E_\gamma(\k) - E_s(\p+\k)
   \simeq
   \frac{p_\perp^2 + m_s^2}{2p}
   + \frac{k_\perp^2 + m_\gamma^2}{2k}
   - \frac{|\p_\perp+\k_\perp|^2 + m_s^2}{2(p+k)} \,,
\label {eq:deltaE}
\end {equation}
where
${\bm P}=\p+\k$ is the original momentum and
the $m$ are the effective finite-temperature masses of the
particles.
The formation time is the quantum mechanical duration of the
off-shell state, $\lf \sim (\delta E)^{-1}$.
If one ignores the masses,
(\ref{eq:deltaE}) can be rewritten in the form
\begin {equation}
   \lf \sim \frac{1}{\delta E}
   \sim \frac{p+k}{p k\bigl(\frac{p_\perp}{p}-\frac{k_\perp}{k}\bigr)^2}
   \simeq \frac{1}{x(1-x)E \theta^2} \,.
\label {eq:lf}
\end {equation}
For $x$ not too close to 1 (i.e.\ $1{-}x$ not small), this is the
same parametric estimate as (\ref{eq:lpm1}).
As stated earlier, I will focus on $x<1/2$ and
ignore the case of small $1{-}x$ in the
main text.

Now, to set some scales, consider QED bremsstrahlung in an infinite
medium, which is dominated by typical scattering events.
The dominant photons are those whose angle $\theta$
relative to the charged particle is of order the net deflection angle
$\Delta\theta$ of the charged particle from the scatterings it
experiences during one formation time: Bremsstrahlung at larger
angles $\theta \gg \Delta\theta$ is suppressed.
On the other hand, because of multiple collisions,
the average angle $\theta$
that the photon makes with the charged particle during a formation
time cannot be smaller than order $\Delta\theta$.  So
(\ref{eq:lpm1}) becomes
\begin {equation}
   \linf \sim \frac{1}{\omega (\Delta \theta)^2} \,.
\label {eq:linfQED1}
\end {equation}
In an infinite medium (or any medium larger than the formation
length),
the typical deflection angle $\Delta\theta$ of the charged particle
in one formation length $\linf$
is
\begin {equation}
   (\Delta\theta)_{\infty} \sim \frac{Q_{\perp\rm\infty}}{E}
                \sim \frac{(\hat q \linf)^{1/2}}{E} \,,
\label {eq:linfQED2}
\end {equation}
where $Q_{\perp\infty} \sim (\hat q l_\infty)^{1/2}$
is the transverse momentum the charged particle
picks up over that distance.  Combining (\ref{eq:linfQED1}) and
(\ref{eq:linfQED2}),
\begin {equation}
   \linf(\omega) \sim \sqrt{ \frac{E^2}{\hat q \omega} } \,.
\label {eq:linfQED}
\end {equation}
Energy loss in an infinite medium is dominated by the case
$\omega \sim E$ where the photon carries away a significant
fraction of the particle's energy.  In this case, the formation
length becomes $L_\infty \sim \sqrt{E/\hat q}$, just like the
QCD result quoted in (\ref{eq:Linf}).

  For the sake of easy reference, and for comparing and contrasting
QED and QCD bremsstrahlung, I have collected in Table
\ref{tab:stuff} some of the formulas described here and in
section \ref{sec:QCD}.

\begin {table}

\begin {tabular}{|l||c|c|l|}
   \multicolumn{1}{l}{~}
   & \multicolumn{1}{c}{QED}
   & \multicolumn{1}{c}{QCD}
   \\
\hline
$\lf$
   & $\frac{1}{\omega\theta^2} \sim \frac{1}{\omega (\Delta\theta)^2}$
   & $\frac{1}{\omega\theta^2} \sim \frac{1}{\omega (\Delta\theta)_{\rm g}^2}$
   & ~formation length (general)
   \\
$\theta$
   & $\Delta\theta \sim \frac{Q_\perp}{E}$
   & $(\Delta\theta)_{\rm g} \sim \frac{Q_\perp}{\omega}$
   & ~characteristic bremsstrahlung angle
   \\
$(Q_\perp^2)_{\rm typ}$ for $L \lesssim \linf$
   & $\hat q L$
   & $\qhatA L$
   & ~typical momentum transfer
   \\
$\linf(\omega)$
   & $\sqrt{ \frac{E^2}{\hat q \omega} }$
   & $\sqrt{\frac{\omega}{\qhatA}}$
   & ~infinite-medium formation length
   \\
$L_\infty \sim l_\infty(E/2)$
   & $\sqrt{ \frac{E}{\hat q} }$
   & $\sqrt{ \frac{E}{\hat q} }$
   & ~dominant $l_\infty$ for energy loss
   \\
$\hat q$
  & $\alpha^2 n \ln\left(\frac{\hat q L}{\md^2}\right)$
  & $\qhatA \sim \ca \alpha^2 {\cal N} \ln\left(\frac{\qhatA L}{\md^2}\right)$
  & ~relevant $Q_\perp^2$ per length
  \\
\hline
\end {tabular}
\caption {
    \label {tab:stuff}
    Summary of various parametric formulas for the LPM effect in
    QED and QCD.  These formulas assume $1{-}x$ is not small, and
    the QCD formulas ignore the difference between $\qhatA$ and
    $\hat q_{\rm F}$ in the case $\omega \sim E$.
}
\end {table}


\subsection {Bremsstrahlung for a given $Q_\perp$}

Let $Q_\perp$ be the total transverse momentum transferred to
the charged particle as it traverses a medium of length $L$,
and consider the case $L \ll \linf(\omega)$ of a medium
that is thin compared to the typical
infinite-volume formation length (\ref{eq:linfQED}) for photons with some
frequency $\omega$.  In this section, I will focus on how
the medium effect on the bremsstrahlung probability depends
on $Q_\perp$.


\subsubsection {LPM suppression in thin media}

Consider a charged, nearly massless particle that undergoes a single
scattering in vacuum.  The probability that this scattering will produce
a photon with frequency of order $\omega$ (for any $\omega \lesssim E$)
is of order $\alpha$ times a collinear logarithm.
For the vacuum case,
integration over bremsstrahlung frequencies
gives rise to an additional, infrared logarithm in the probability.  In
considering medium effects on bremsstrahlung, I am going to delay
integration over frequency until the end and will for now just consider
the probability of emission of photons whose frequencies are of order
some scale $\omega$.  The collinear logarithm comes from integrating
over photon directions that are very close to either the incoming or
outgoing particle track in an isolated collision.
It plays only a limited role in
medium effects on bremsstrahlung.
For the moment, I will ignore the collinear logarithm and simply take
$\alpha$ to be the additional cost in probability of emitting a photon of
frequency $\sim \omega$ when there is a collision.  The angle this
photon makes with the incoming and outgoing particle tracks
is of order the deflection angle of the particle track,
with discussion of the possibility of more-nearly collinear photons
deferred to later discussion of the collinear logarithm.

In a medium,
divide the multiple collisions along a trajectory into
sets which are each roughly one formation length long.
The photon cannot resolve the difference of a single vs.\ multiple
collision in each set, but it does see each set as a distinct
opportunity for bremsstrahlung.
There is then an $O(\alpha)$ probability for a photon emission
from each set for which $\Delta\theta \sim \theta$.
So, for instance, the shaded oval
in Fig.\ \ref{fig:NRrare}a is associated with a probability of
$O(\alpha)$ [times a collinear logarithm] for emitting a
bremsstrahlung photon at the angle shown.

\begin {figure}
\begin {center}
  \includegraphics[scale=0.6]{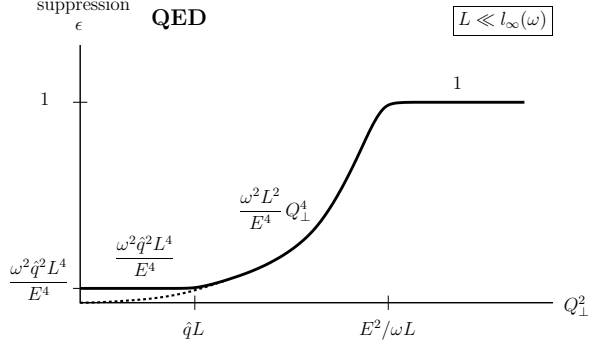}
  \caption{
     \label{fig:qedLPM}
     Suppression factor $\epsilon$ of medium effects
     as a function of $Q_\perp^2$ for the
     case where the medium thickness $L$ is smaller
     than the typical formation length $\linf(\omega)$ of the
     infinite-medium case.
  }
\end {center}
\end {figure}

In Fig.\ \ref{fig:NRboost1}a, there is also an $O(\alpha)$ probability of
emitting such a photon at the angle $\theta \sim \Delta\theta$ shown
there.  But the {\it difference}\/ in emission
probability with the vacuum case of Fig.\ \ref{fig:NRboost1}b is small.
In this case, I will write the medium contribution to the
$O(\alpha)$ probability of photon emission as $O(\epsilon\alpha)$,
where $\epsilon$ is an LPM suppression factor due to the photon's
failure to resolve (i) the collisions in the medium from (ii) the event
that originally created the charged particle:
\begin {equation}
   \Delta(\mbox{cost of bremsstrahlung emission}) \sim \epsilon \alpha .
\label{eq:epsdef}
\end {equation}
Fig.\ \ref{fig:qedLPM} shows the behavior of $\epsilon$ as
a function of $Q_\perp^2$.  I will explain this figure one feature
at a time.

For a given scattering trajectory---{\it e.g.} the typical
scattering events of Fig.\ \ref{fig:NRboost1}a or the
rare events of \ref{fig:NRrare}a, or something in
between---the relevant formation length will depend on the deflection
angle, which for thin media will be related to the net
transverse momentum transfer $Q_\perp$ while traversing the
medium by
\begin {equation}
   \Delta\theta \sim \frac{Q_\perp}{E} .
\label {eq:dtheta1}
\end {equation}
The corresponding formation length (\ref{eq:lpm1}) is
\begin {equation}
   \lf \sim \frac{1}{\omega (\Delta\theta)^2}
   \sim \frac{E^2}{\omega Q_\perp^2} \,.
\label {eq:lfQED}
\end {equation}

Note that in a rare scattering case like Fig.\ \ref{fig:NRrare}a,
I should have estimated the formation length based on the net deflection
over the formation length, shown by the shaded oval, rather then
from the deflection over the entire trajectory.
However, in this case, the rare scattering dominated the total
angular deflection in any case, and so I do not need to
distinguish between the two.

The region $\epsilon \simeq 1$ of Fig.\ \ref{fig:qedLPM}
corresponds to cases $\lf \ll L$
where the scattering process looks like Fig.\ \ref{fig:NRrare}a
rather than Fig.\ \ref{fig:NRboost1}a.
From (\ref{eq:lfQED}), this condition is
equivalent to
\begin {equation}
   Q_\perp^2 \gg \frac{E^2}{\omega L}
   \qquad
   \mbox{for $\epsilon \simeq 1$}.
\end {equation}   


\subsubsection {The size of $\epsilon$}

Now turn to the case $Q_\perp^2 \ll E^2/\omega L$ where the formation
time $\lf$ of (\ref{eq:lfQED}) is large compared to $L$,
corresponding to situations like Fig.\ \ref{fig:NRboost1}a.
Recall that the LPM effect occurs when the relative
phase $K_\mu (X_1-X_2)^\mu$ is small for space-time points
$X_1$ and $X_2$ corresponding to collisions.  For collisions
spread out over a distance $L$ as in Fig.\ \ref{fig:NRboost1}a,
this relative phase is of order
$\omega L (1-\cos\theta) \sim \omega L \theta^2$.
We have LPM suppression if $\omega L \theta^2 \ll 1$, and the
{\it amount}\/ of LPM suppression of the effects of collisions
with the medium turns out to be given by the
square of this relative phase:
\begin {equation}
   \epsilon \sim (K\cdot\Delta X)^2 \sim (\omega L \theta^2)^2
            \sim \left( \frac{L}{\lf} \right)^2 .
\label {eq:LPMsupp}
\end {equation}
I give a brief review in Appendix \ref{app:LPMsupp} of why
this is the amount of suppression.
Putting the formation length (\ref{eq:lfQED})
into (\ref{eq:LPMsupp}),
\begin {equation}
   \epsilon \sim \frac{\omega^2 L^2}{E^4} \, Q_\perp^4 ,
\label {eq:qedLPMsupp}
\end {equation}
as depicted by the fall-off of $\epsilon$ with decreasing
$Q_\perp$ shown in Fig.\ \ref{fig:qedLPM}.

\begin {figure}
\begin {center}
  \includegraphics[scale=0.3]{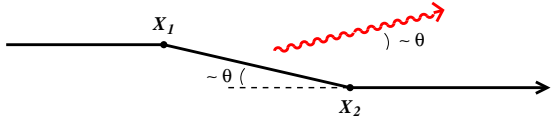}
  \caption{
     \label{fig:smallQT}
     Example of two scatterings with zero total deflection angle.
  }
\end {center}
\end {figure}

For $Q_\perp^2$ small compared to the typical transfer
of $(Q_\perp^2)_{\rm typ} \sim \hat q L$, the solid line
in Fig.\ \ref{fig:qedLPM} deviates from (\ref{eq:qedLPMsupp}),
with the latter indicated by a dotted line.
This qualitative difference will not matter to the eventual conclusions of
this paper, but I will take a moment to explain it for
the sake of completeness.
In earlier discussion, I slightly oversimplified when asserting that
bremsstrahlung
is suppressed if the photon angle $\theta$ is large compared to the
net deflection angle $\Delta\theta \sim Q_\perp/E$ of the charged
particle.  Consider the trajectory shown in Fig.\ \ref{fig:smallQT}.
Here, the net deflection of the trajectory is zero but the intermediate
deflection is non-zero.  The typical bremsstrahlung photon
angle is then of order the intermediate deflection angle, which I show
in more detail in Appendix \ref{app:smallQT}.
Now consider cases of many multiple scatterings within a formation
time.  The rare cases where $N$ multiple scatterings
 produce smaller-than-typical
deflections are dominated by situations where
the first
$N/2$ scatterings produce a typical deflection,
which by random chance was nearly canceled by an
opposite deflection from the second $N/2$
scatterings.%
\footnote{
  For example,
  consider a simple one-dimensional random walk.
  The average displacement after $N$ steps grows as $N^{1/2}$.
  If you look at the small subset of random walks which have
  zero displacement after $N$ steps, their average displacement
  after half of those steps still grows as $N^{1/2}$.
}
The dominant photon angle $\theta$ will then be determined by
the intermediate deflection $\sim (Q_\perp)_{\rm typ}/E$ rather
then the total deflection $\sim Q_\perp/E$ that was used in
(\ref{eq:dtheta1}) and (\ref{eq:lfQED}).  The upshot is that
$Q_\perp \ll (Q_\perp)_{\rm typ}$ collisions behave just like
$Q_\perp \sim (Q_\perp)_{\rm typ}$ collisions in terms of
the medium effect on bremsstrahlung.

\subsubsection {Putting it together}

\begin {figure}
\begin {center}
  \includegraphics[scale=0.6]{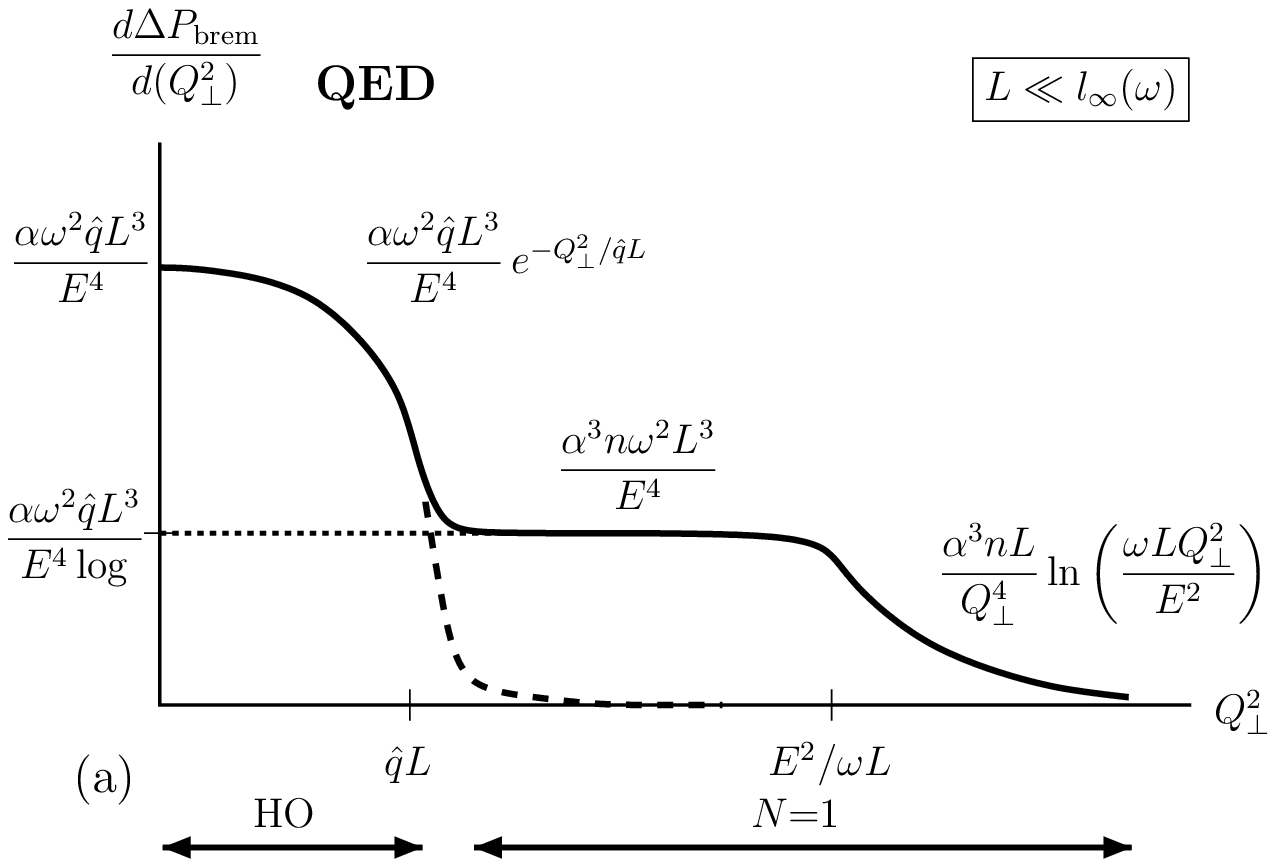}
  \hspace{0.1in}
  \includegraphics[scale=0.6]{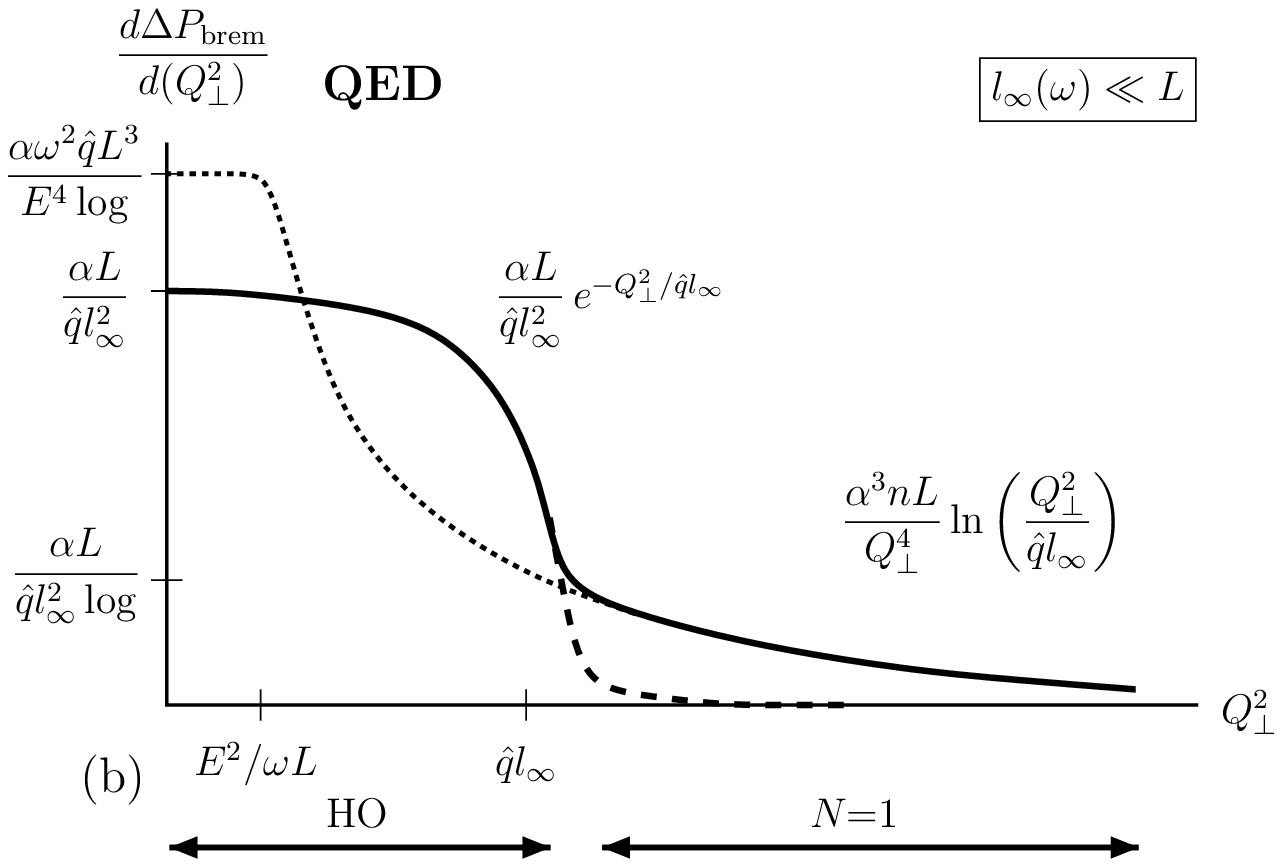}
  \caption{
     \label{fig:qedPbrem}
     Medium modification to probability 
     for emitting a high-energy bremsstrahlung gluon of
     frequency $\sim\omega$: probability distribution vs.\ $Q_\perp^2$.
     In (b), $Q_\perp$ represents the transverse momentum picked
     up in a distance $l_\infty(\omega)$.
  }
\end {center}
\end {figure}

For a thin medium,
we can now find a parametric result for the medium effect
$\Delta P_{\rm brem}$ on the
probability of single-photon bremsstrahlung for photons of
frequency $\sim\omega$ through scattering
processes with total momentum transfer $\sim Q_\perp$.
Neglecting collinear logarithms, it is simply the probability of
the underlying scattering event times a factor of $\epsilon \alpha$
for the associated bremsstrahlung.  Multiplying
(i) Fig.\ \ref{fig:Pscatt} for $dP_{\rm scatt}/d(Q_\perp^2)$
times (ii) Fig.\ \ref{fig:qedLPM} for $\epsilon$ times (iii)
$\alpha$ gives
\begin {equation}
   \frac{d(\Delta P_{\rm brem})}{d(Q_\perp^2)}
   \sim \frac{dP_{\rm scatt}}{d(Q_\perp^2)}
        \times \epsilon(\omega,Q_\perp^2) \times \alpha ,
\label {eq:product}
\end {equation}
which is depicted in Fig.\ \ref{fig:qedPbrem}a.
There is an additional logarithmic factor shown for
the high-$Q_\perp$ tail in
Fig.\ \ref{fig:qedPbrem}a that is not included in the
product (\ref{eq:product}).  This is a collinear logarithm that
I will explain in a moment.  Recall that the notation
$\Delta P_{\rm brem}$ is defined in terms of the
frequency spectrum by (\ref{eq:DPbremdef}).

Fig.\ \ref{fig:qedPbrem}b shows the corresponding result for
a thick medium $L \gg l_\infty(\omega)$.
Since relevant formation lengths in this case will not exceed
$\linf(\omega)$, we can break the problem up into independent
probabilities for each section of medium of length $\linf$.
For a section of length $\linf$, we have $\epsilon \sim 1$.
So (\ref{eq:product}) is modified to%
\footnote{
  The dotted line in Fig.\ \ref{fig:qedPbrem}b
  showing the $N{=}1$ result for $Q_\perp^2 \ll \hat q \, l_\infty(\omega)$
  is determined by (\ref{eq:product}) instead of (\ref{eq:product2}).
  In the $N{=}1$ approximation, $\lf$ can exceed $\linf(\omega)$ for small
  total deflection angle $Q_\perp/E$.
}
\begin {equation}
   \frac{d(\Delta P_{\rm brem})}{d(Q_\perp^2)}
   \sim \frac{L}{\linf(\omega)} \times
        \left[ \frac{dP_{\rm scatt}}{d(Q_\perp^2)} \right]_{L=\linf(\omega)}
        \times \alpha ,
\label {eq:product2}
\end {equation}
where here $Q_\perp$ refers to the transverse momentum transfer over
a length of order $\linf(\omega)$.
The parametric behaviors shown in Figs.\ \ref{fig:qedPbrem}a and
b agree for the dividing case of $L \sim \linf(\omega)$.

When discussing ``thick'' media, I have implicitly assumed that
a single bremsstrahlung analysis of the medium effect remains adequate.
In particular,
the media should be small compared to the stopping
distance for the high-energy particle: $L \ll L_\infty/\alpha$.
The stopping distance is where $\Delta E \sim E$, as can be read
off from Fig.\ \ref{fig:qcdEL} for QCD or later from Fig.\
\ref{fig:qedEL} for QED.


\subsubsection {Collinear logarithms}
\label {sec:collinear}

\begin {figure}
\begin {center}
  \includegraphics[scale=0.3]{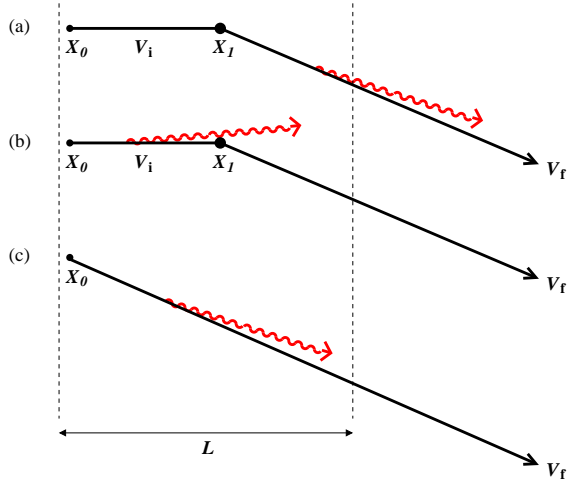}
  \caption{
     \label{fig:collinear}
     (a) initial and (b) final state collinear radiation from
     a single scattering with the medium.  For comparison,
     (c) shows the case
     of collinear radiation when there are no medium collisions
     but only the initial hard scattering event that created
     the high-energy particle moving in the same final direction.
  }
\end {center}
\end {figure}

I will now discuss collinear logarithms associated
with bremsstrahlung, which were ignored in the previous analysis.
For simplicity, consider the case $\omega \ll E$ of soft bremsstrahlung,
where the charged particle trajectory can be thought of as
a classical source for the electromagnetic field.
For further simplicity, start by considering a trajectory
corresponding to exactly one scattering from the medium, as in
Fig.\ \ref{fig:NRrare}c, rather than more complicated trajectories
like Figs.\ \ref{fig:NRrare}a or \ref{fig:NRboost1}a that include multiple
small-angle scatterings.
In this case, there is a collinear logarithm in the bremsstrahlung
probability associated with small photon angles
$\theta \ll \Delta\theta$ relative to the final particle
direction, as in Fig.\ \ref{fig:collinear}a.
There is also potentially a similar logarithm
associated with collinearity with the initial particle direction,
as in Fig.\ \ref{fig:collinear}b,
but this second logarithm can be suppressed by the LPM effect.

Throughout this discussion, I will assume that $E$ and $\omega$ are
large enough compared to effective masses that I can treat the
charged particle and photon as massless.  So I will not keep track
of the cut-off of collinear logarithms due to masses.

In the case of final-state collinearity shown in
Fig.\ \ref{fig:collinear}a, the enhancement of the bremsstrahlung
probability at small angles is the same as that
for the vacuum process of Fig.\ \ref{fig:collinear}c.
There is therefore no corresponding collinear logarithm in the
medium effect $\Delta P_{\rm brem}$, which expresses the difference
between the two.  More detail is given in Appendix \ref{app:collinear}.

Now consider collinearity with the earlier direction of
the particle trajectory, as in Fig.\ \ref{fig:collinear}b,
and let $\theta_{{\rm i}\gamma}$ be the small angle that the photon
makes with that direction.
The corresponding collinear logarithm will be cut off at small
$\theta_{{\rm i}\gamma}$
when the formation length (\ref{eq:lfQED}) becomes large compared
to the length of that segment of the trajectory, because then
the photon cannot resolve the difference between
the particle trajectories of Figs.\ \ref{fig:collinear}b and
\ref{fig:collinear}c.  A generic collision in the medium will
be a distance of order $L$ from the start, and so the angles
which contribute to a collinear logarithm must satisfy
$\lf(\theta_{{\rm i}\gamma}) \lesssim L$, which is
\begin {equation}
   \frac{1}{\omega \theta_{{\rm i}\gamma}^2}
   \lesssim L .
\end {equation}
The angles which contribute to the collinear logarithm
are therefore
\begin {equation}
   \sqrt{ \frac{1}{L \omega} }
   \lesssim \theta_{{\rm i}\gamma}
   \lesssim \Delta\theta .
\label {eq:thetai}
\end {equation}
The collinear logarithm appears when there exists such a hierarchy
of angular scales, and it is then
\begin {equation}
   \ln \left(
     \frac{(\Delta\theta)^2}{(\theta_{{\rm i}\gamma})_{\rm min}^2}
   \right)
   \sim \ln\left( \frac{\omega L Q_\perp^2}{E^2} \right) .
\end {equation}
This is the logarithmic factor shown on the large $Q_\perp^2$ tail
of Fig.\ \ref{fig:qedPbrem}a.
The range (\ref{eq:thetai}) only exists if
$Q_\perp^2 \gg E^2/\omega L$.

So far, I have considered single scattering processes like
Fig.\ \ref{fig:collinear}a--b rather than actual
cases of interest to this paper, such as Fig.\ \ref{fig:NRrare}a.
The angle that the photon makes with the trajectory preceding
the relatively large angle collision in Fig.\ \ref{fig:NRrare}a
is smeared out by multiple soft scatterings,
which deflected the particle by an angle of order
$(Q_\perp)_{\rm typ}/E \sim \sqrt{\hat q L/E^2}$.
The angular
range (\ref{eq:thetai}) contributing to a collinear logarithm
is then replaced by
\begin {equation}
   \max\left(
      \sqrt{ \frac{1}{\omega L} } \,,
      \sqrt{ \frac{\hat q L}{E^2} } 
   \right)
   \lesssim \theta_{{\rm i}\gamma}
   \lesssim \Delta\theta .
\label {eq:thetai2}
\end {equation} 
The first case on the left-hand side dominates when
$L \ll \linf(\omega)$, as in Fig.\ \ref{fig:qedPbrem}a.
For $L \gg \linf(\omega)$, the relevant length scale is
$\linf(\omega)$ rather than $L$, and the logarithm becomes
\begin {equation}
   \ln \left(
     \frac{(\Delta\theta)^2}{(\theta_{{\rm i}\gamma})_{\rm min}^2}
   \right)
   \sim \ln\left( \frac{Q_\perp^2}{\hat q \, \linf(\omega)} \right) .
\label {eq:thicklog}
\end {equation}
This is argued in more detail in Appendix \ref{app:thicklog}.
The logarithmic factor (\ref{eq:thicklog}) is the one
shown on the large $Q_\perp^2$ tail
of Fig.\ \ref{fig:qedPbrem}b.

Now return to the case of final-state radiation in a case
like Fig.\ \ref{fig:NRrare}a.
Once the particle leaves the medium, there is always a
semi-infinite straight line segment of the trajectory to which
a photon can become collinear.  So, unlike the case just considered,
the effect of multiple soft scatterings before the particle
leaves the medium cannot suppress the production of
final-state collinear photons: photons can be produced at
arbitrarily small angles (if the charged particle is
treated as massless) by being produced after the very last
scattering in the medium.  But this contribution cancels when
we subtract the vacuum contribution to get $\Delta P_{\rm brem}$,
just as discussed earlier for the case of Fig.\ \ref{fig:collinear}a.


\subsection {Bremsstrahlung spectrum and energy loss}

\begin {figure}
\begin {center}
  \includegraphics[scale=0.6]{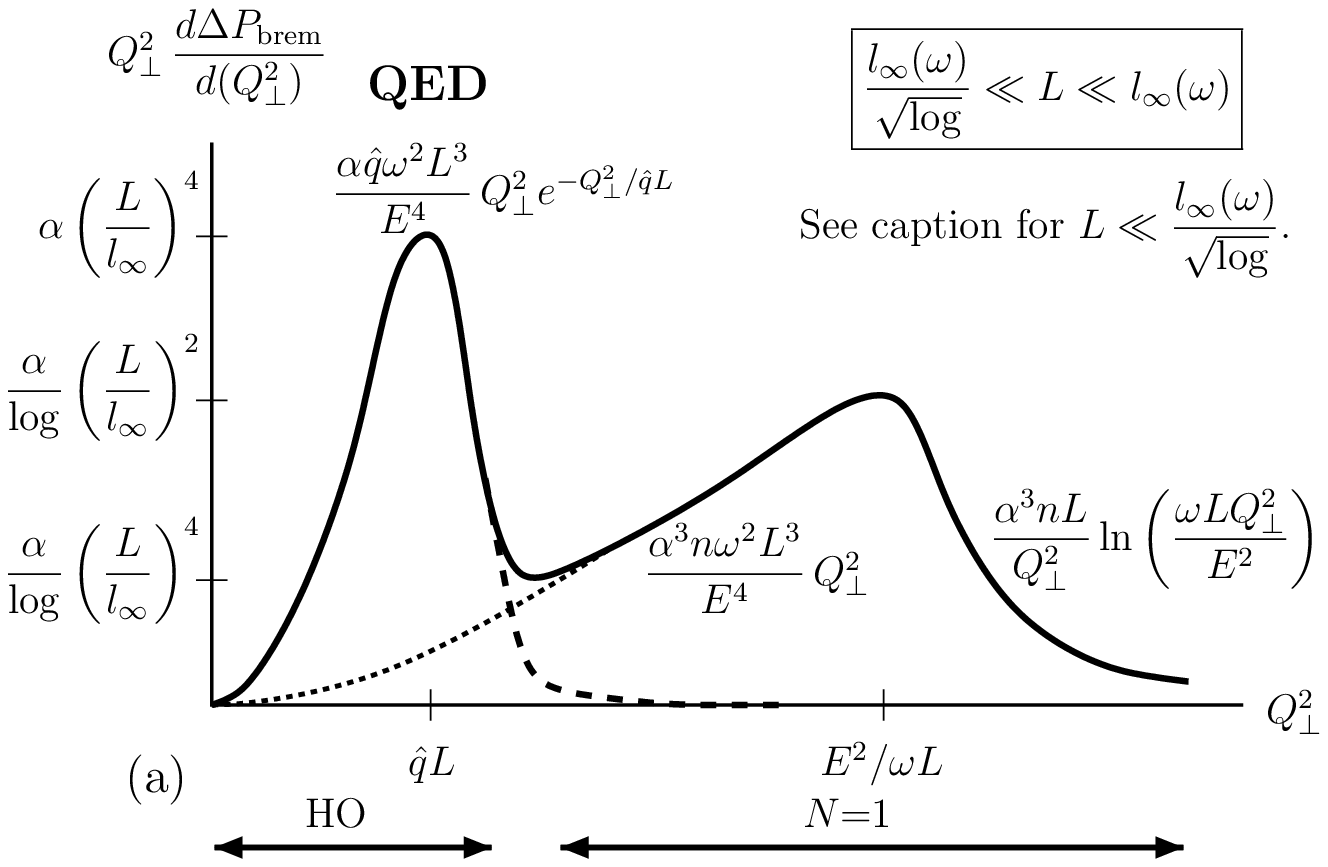}
  \hspace{0.03in}
  \includegraphics[scale=0.6]{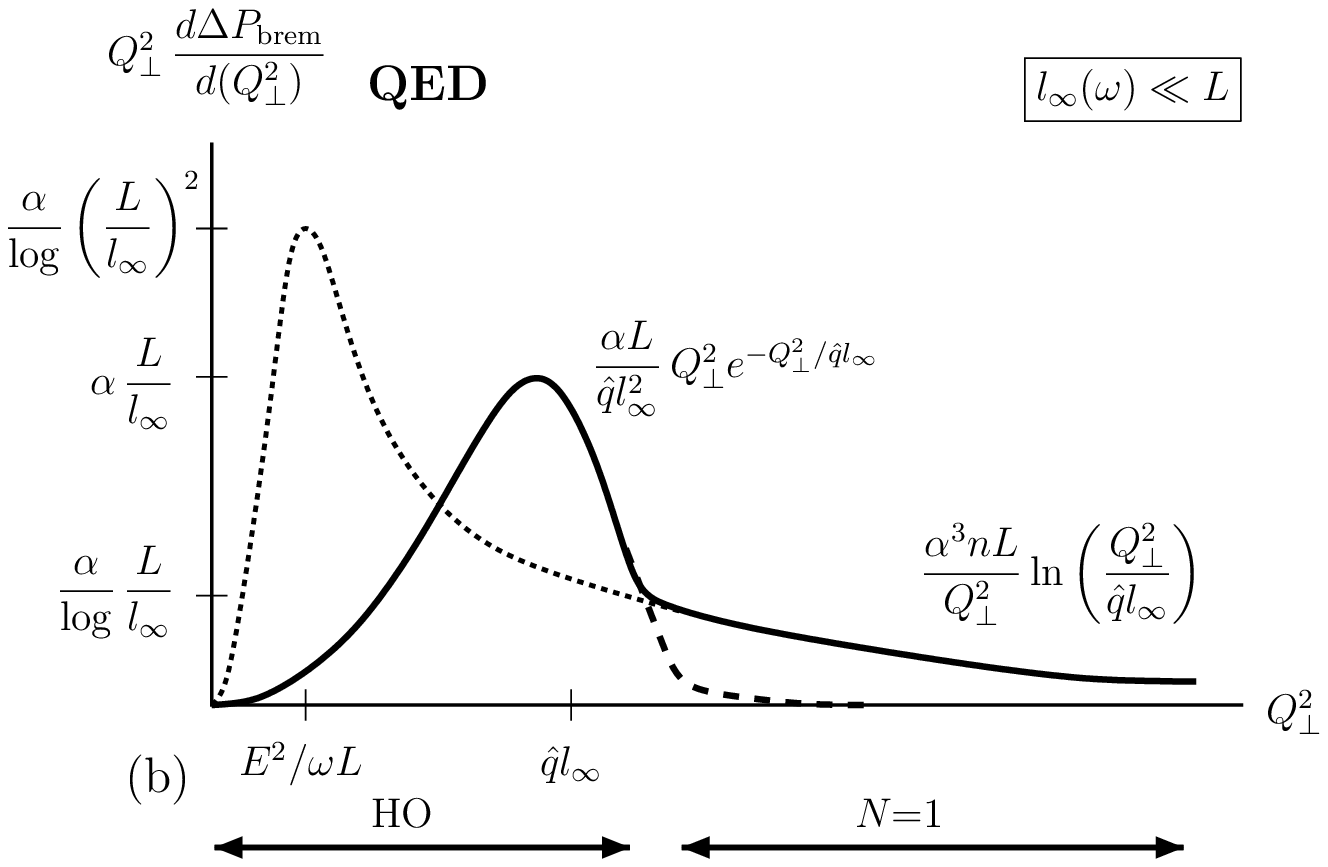}
  \caption{
      \label{fig:qedPbremQ}
      This is Fig.\ \ref{fig:qedPbrem} multiplied by $Q_\perp^2$.
           (a) assumes that
           $l_\infty(\omega)/\sqrt{\log} \ll L \ll l_\infty(\omega)$;
           the figure is similar
           for $L \ll l_\infty(\omega)/\sqrt{\log}$, but then
           the right-hand ($N{=}1$) peak would
           be the higher one.
     }
\end {center}
\end {figure}

I now want to integrate over $Q_\perp^2$ to find the spectrum
$\Delta P_{\rm brem} = \omega \> d(I-I_{\rm vac})/d\omega$
as a function of $\omega$.
To see visually what values of $Q_\perp$ dominate the integration,
it is useful to multiply the previous results of
Fig.\ \ref{fig:qedPbrem} by a factor of $Q_\perp^2$ so that
$d(\Delta P_{\rm brem})/d(Q_\perp^2)$ becomes the logarithmic derivative
$Q_\perp^2 \> d(\Delta P_{\rm brem})/d(Q_\perp^2)$.  The result is shown in
Fig.\ \ref{fig:qedPbremQ}.  In the thin-media case, there are
two peaks, corresponding to two different $Q_\perp$ scales that
will give the dominant contributions to the integral.
One scale is the scale $Q_\perp^2 \sim \hat q L$ of typical
scattering events, corresponding to HO processes like
Fig.\ \ref{fig:NRboost1}a.  The formation length $\lf$ in this case
is large compared to $L$.
The other, larger $Q_\perp$ scale
corresponds to rarer, larger-angle scattering events that are well
approximated by the $N=1$ approximation and which have a
shorter formation length than the typical scattering events.
The scale $Q_\perp^2 \sim E^2/\omega L$ of the right-hand peak
in Fig.\ \ref{fig:qedPbremQ}a corresponds to the case where this
shorter formation length is of order $L$.  The tail at yet larger
$Q_\perp$ corresponds to yet shorter formation lengths, as were
depicted in Fig.\ \ref{fig:NRrare}.  This tail falls off
because no further gains are made in the suppression factor
$\epsilon$ by further decreasing $\lf$
below $L$, but the probability of the underlying scattering
event decreases.

Which of the peaks of Fig.\ \ref{fig:qedPbremQ}a
dominates in the thin-media case of
$L \ll \linf(\omega)$ depends on the exactly how small $L$ is.
The HO and $N{=}1$ peak heights are
\begin {equation}
   \Delta P_{\rm brem}(\omega) 
   \sim
   \frac{\alpha \hat q^2 \omega^2 L^4}{E^4}
   \sim
   \alpha \left( \frac{L}{\linf(\omega)}\right)^4
   \hspace{7em}
   (\mbox{HO})
\label {eq:HOpeak}
\end {equation}
and
\begin {equation}
   \Delta P_{\rm brem}(\omega)
   \sim
   \frac{\alpha^3 n \omega L^2}{E^2}
   \sim
   \frac{\alpha}{\ln(\hat q L / \md^2)}
   \left( \frac{L}{\linf(\omega)}\right)^2
   \hspace{4em}
   (N{=}1)
\label {eq:N1peak}
\end {equation}
respectively.
The $N{=}1$ peak dominates when
\begin {equation}
   L \ll
   \frac{\linf(\omega)}{\bigl[\ln(\hat q L / \md^2)\bigr]^{1/2}} \,.
\end {equation}

The $Q_\perp$-integrated result $\Delta P_{\rm brem}$ is of order
the highest peak in Fig.\ \ref{fig:qedPbremQ}.  The result is
shown vs.\ $L$ in Fig.\ \ref{fig:qcdPbremL}, provided one takes
the QED formula (\ref{eq:linfQED}) for
$\linf = \linf(\omega)$ instead of (\ref{eq:linfQCD}).

\begin {figure}
\begin {center}
  \includegraphics[scale=0.6]{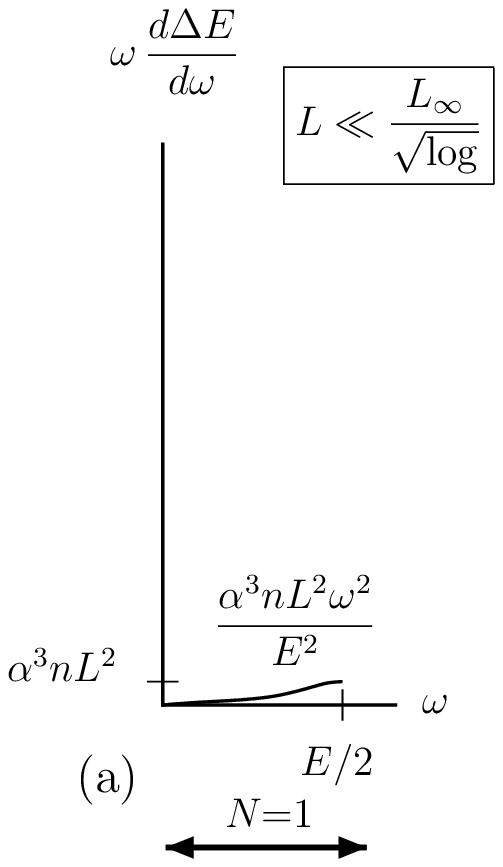}
  \hspace{0.1in}
  \includegraphics[scale=0.6]{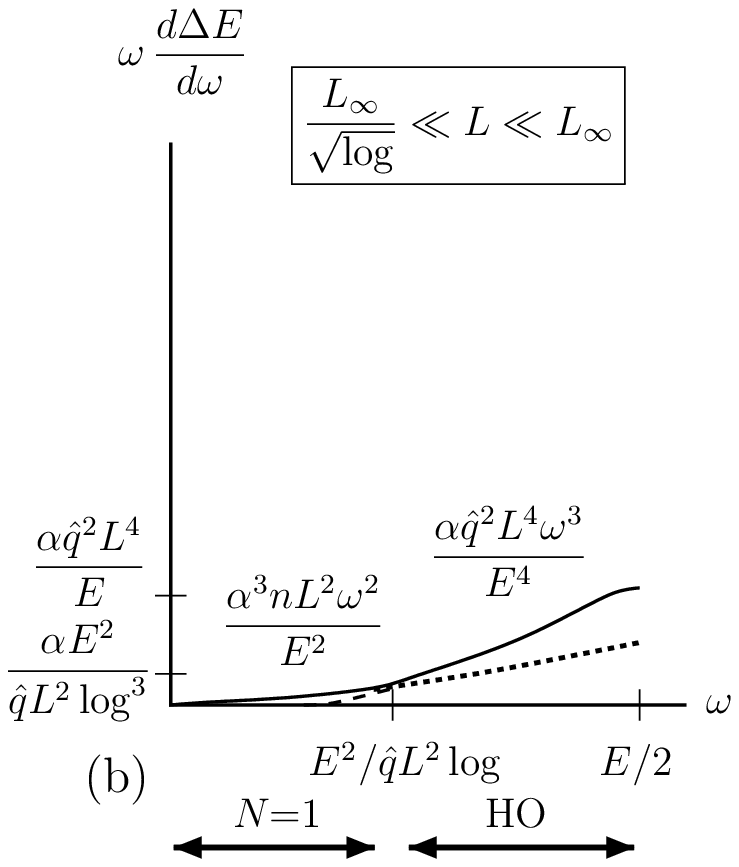}
  \hspace{0.1in}
  \includegraphics[scale=0.6]{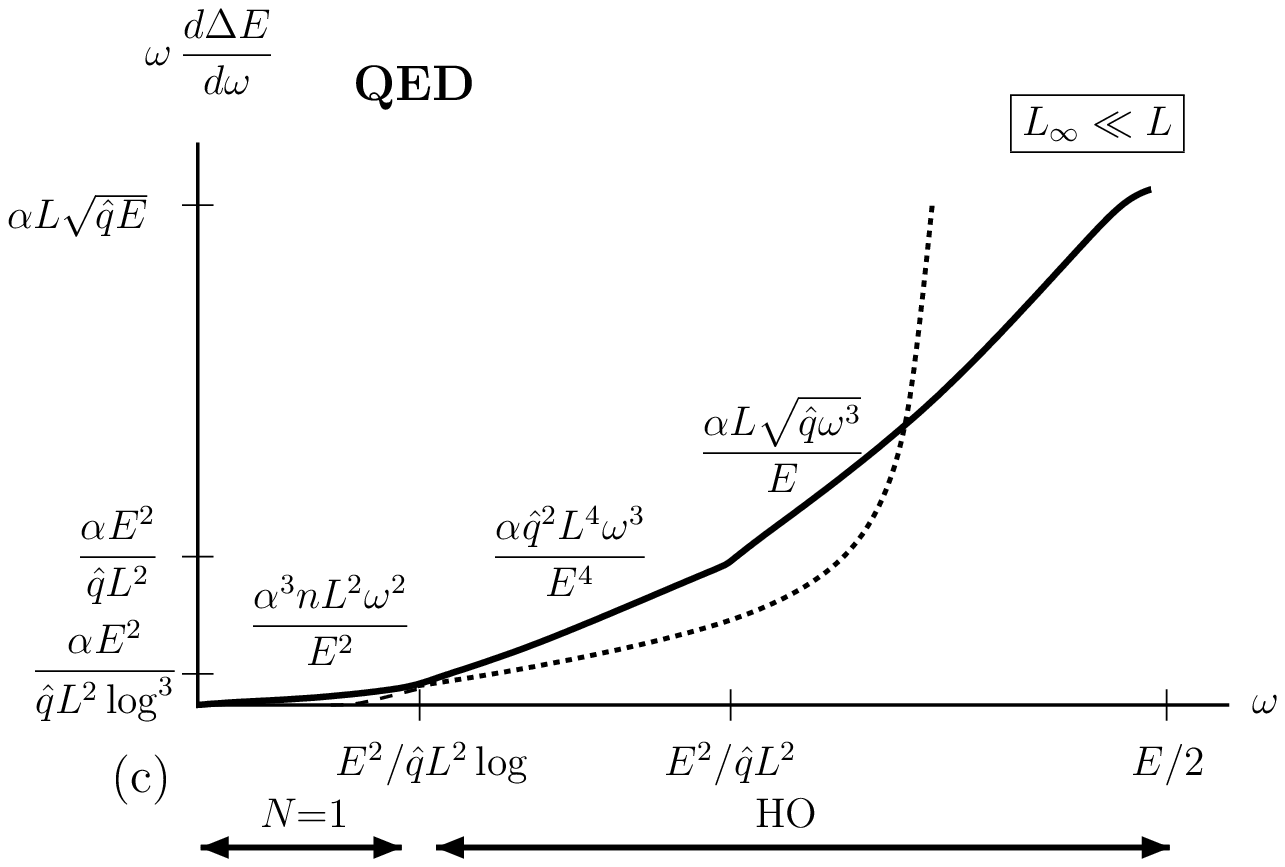}
  \caption{
     \label {fig:qedE}
     Medium modification to QED energy loss $\Delta E$:
           contribution of bremsstrahlung photons with
           frequency $\sim \omega$.
  }
\end {center}
\end {figure}

Now consider the dependence on $\omega$ when $L$ is fixed.
This behavior is shown in Fig.\ \ref{fig:qedE} except that
I multiply by an extra factor of $\omega$ to
plot the contribution
\begin {equation}
   \omega \, \frac{d(\Delta E)}{d\omega}
   = \omega^2 \, \frac{d}{d\omega}(I-I_{\rm vac})
   = \omega \, \Delta P_{\rm brem}(\omega)
\end {equation}
to the medium effect on average energy loss from photons with frequency
$\sim\omega$.

Fig.\ \ref{fig:qedE}c shows the case for thick media $L \gg L_\infty$,
where $L_\infty$ is given by (\ref{eq:Linf}) and represents the
typical formation length for the case $\omega \sim E$.  In QED
(unlike QCD), the typical formation length $\linf(\omega)$
given by (\ref{eq:linfQED}) grows with decreasing $\omega$.
For $\omega$ relatively large, $\linf(\omega)$ will still exceed $L$,
and $\Delta P_{\rm brem}$ will be given by the peak height
\begin {equation}
   \Delta P_{\rm brem}(\omega) \sim \alpha \, \frac{L}{\linf(\omega)}
   \sim \frac{ \alpha L \sqrt{\hat q \omega} }{E}
\end {equation}
of Fig.\ \ref{fig:qedPbremQ}b.
When multiplied by $\omega$, this gives the corresponding formula
shown on the right of Fig.\ \ref{fig:qedE}c.
This formula works until $\omega$ gets
small enough that $\linf(\omega) \sim L$, which occurs at
$\omega \sim E^2/\hat q L^2$.
For smaller $\omega$, first the HO peak and then the
$N{=}1$ peak of Fig.\ \ref{fig:qedPbremQ}a will determine
$\Delta P_{\rm brem}$, giving
(\ref{eq:HOpeak}) and (\ref{eq:N1peak}) respectively,
corresponding to the other formulas shown in Fig.\ \ref{fig:qedE}c.

In the case of $L \ll L_\infty$, shown in Figs.\ \ref{fig:qedE}a and b,
we always have $\linf(\omega) \gg L$, and one or more of the stages
just described are bypassed.

\begin {figure}
\begin {center}
  \includegraphics[scale=0.6]{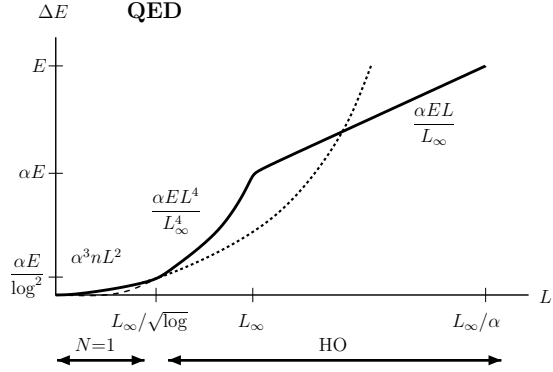}
  \caption{
     \label {fig:qedEL}
     Total medium modification to QED energy loss $\Delta E$,
     shown vs.\ medium length $L$.
  }
\end {center}
\end {figure}

Finally, integrating over $\omega$, the total medium contribution
$\Delta E$ to average energy loss just corresponds parametrically
to the maximum in Fig.\ \ref{fig:qedE}.  A sketch of
the resulting dependence of $\Delta E$ on medium thickness $L$ is
shown in Fig.\ \ref{fig:qedEL}.  This result is qualitatively
different from the QCD result previewed in
Fig.\ \ref{fig:qcdEL}, for reasons which will be explained
in the next section.
The $L^2$ dependence of the QED result for sufficiently small $L$ has been
discussed previously by Peign\'e and Smilga \cite{PeigneSmilga}.

\section{Bremsstrahlung in QCD}
\label {sec:QCD}

\subsection {Review of LPM Effect in QCD}

The major difference between bremsstrahlung in QCD and QED is that
a bremsstrahlung gluon carries charge and so, like the particle that
radiated, can also $t$-channel scatter from the medium.  It is easier
to deflect a lower momentum particle than a higher momentum particle
and so, for the case $\omega \ll E$, it is scattering of the gluon
rather than the original particle that dominates determination of
the angle $\theta$ between the two.  The medium effect on
bremsstrahlung is therefore
dominated by
\begin {equation}
   \theta \sim (\Delta\theta)_{\rm g} \sim \frac{Q_\perp}{\omega} \,,
\label {eq:dtheta2}
\end {equation}
instead of the corresponding QED angle (\ref{eq:dtheta1}).
Correspondingly, here $Q_\perp$ is the transverse momentum that the
bremsstrahlung gluon, rather than the original particle,
picks up in a formation time.
For the case $\omega \sim E$, the deflections (\ref{eq:dtheta2}) and
(\ref{eq:dtheta1}) of the gluon and the original particle are
parametrically the same size, up to details of group Casimirs
related to whose $Q_\perp$ we consider, which I will not bother
to distinguish in my parametric estimates in this case.  So I will
use (\ref{eq:dtheta2}) for the entire range $\omega \lesssim E$,
assuming as always that $1{-}x$ is not small.

The formation length corresponding to (\ref{eq:lpm1}) and
(\ref{eq:lfQED}) is then
\begin {equation}
   \lf \sim \frac{1}{\omega (\Delta\theta)_{\rm g}^2}
   \sim \frac{\omega}{Q_\perp^2} \,.
\label {eq:lfQCD}
\end {equation}
The argument for the size of the formation length $\linf(\omega)$
in an infinite medium then goes through just as in
(\ref{eq:linfQED1}--\ref{eq:linfQED}) for the QED case, but with
$\Delta\theta$ and $E$ replaced by $(\Delta\theta)_{\rm g}$ and
$\omega$, so that
\begin {equation}
   \linf(\omega) \sim \sqrt{\frac{\omega}{\qhatA}} \,,
\label {eq:linfQCD2}
\end {equation}
as quoted earlier in (\ref{eq:linfQCD}).
The main qualitative difference between QCD and QED bremsstrahlung
in a medium is that the QCD formation length $\linf(\omega)$ {\it decreases}
in the soft limit of decreasing $\omega$ due to the ease with
which a soft gluon is deflected, whereas in QED $\linf(\omega)$
increases with decreasing $\omega$.


\subsection {Bremsstrahlung spectrum and energy loss}

\begin {figure}
\begin {center}
  \includegraphics[scale=0.6]{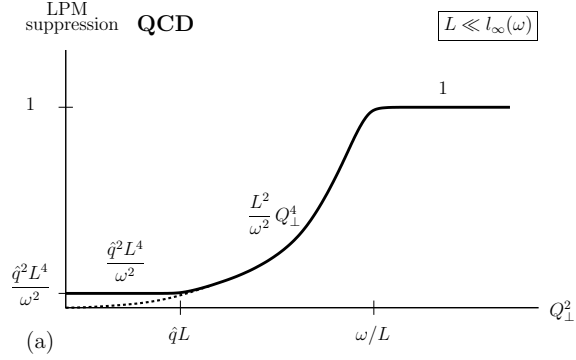}
  \caption{
     \label{fig:qcdLPM}
     The QCD analog of Fig.\ \ref{fig:qedLPM} for the suppression
     factor $\epsilon$.
   }
\end {center}
\end {figure}

\begin {figure}
\begin {center}
  \includegraphics[scale=0.6]{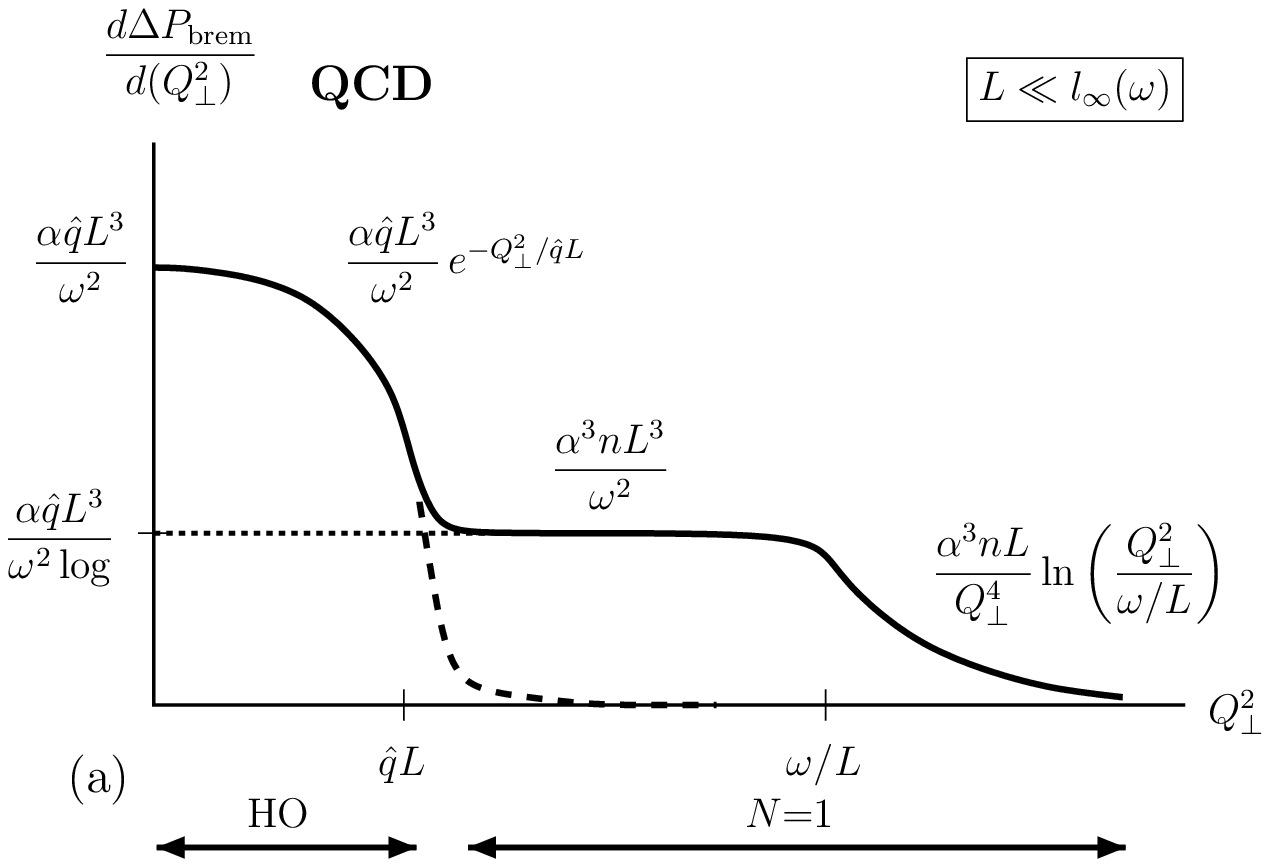}
  \hspace{0.1in}
  \includegraphics[scale=0.6]{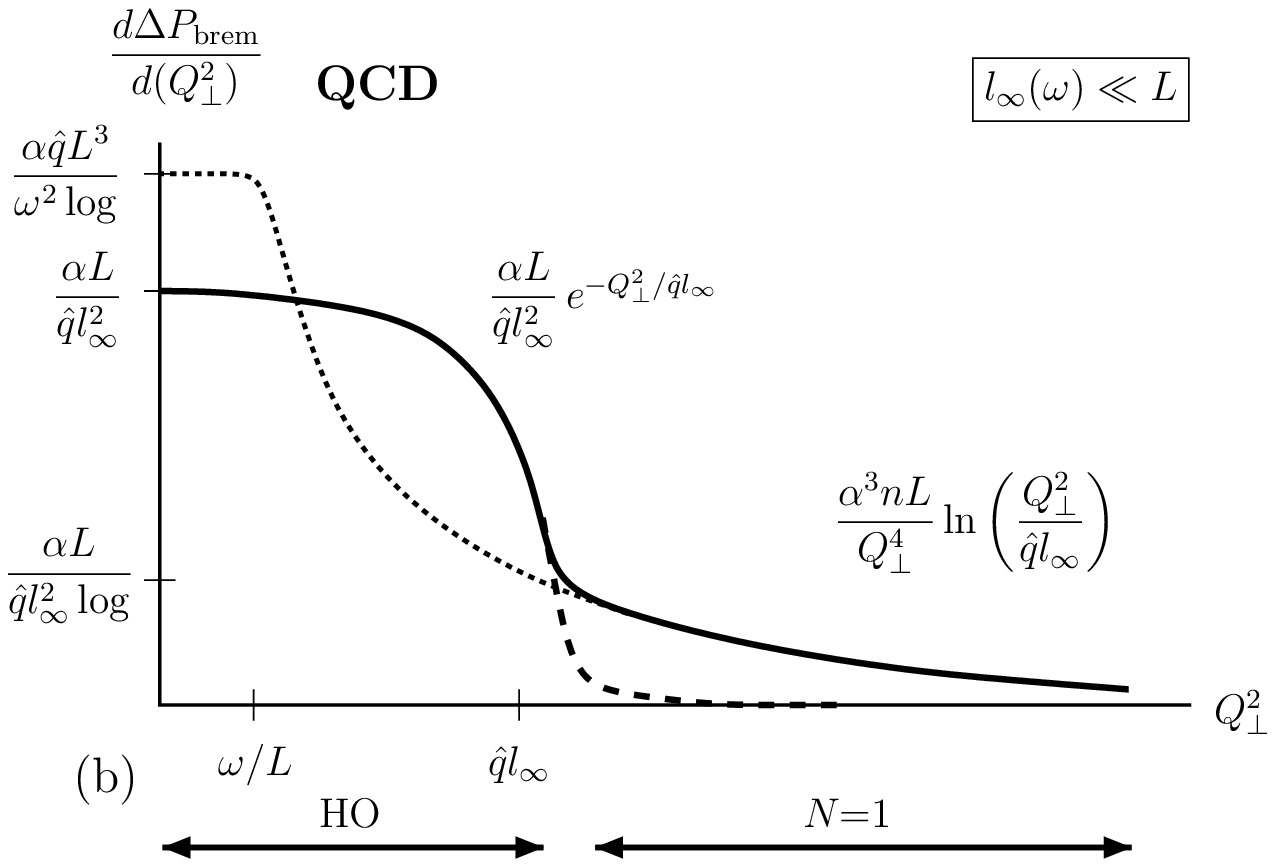}
  \caption{
     \label {fig:qcdPbrem}
     The QCD analog of Fig.\ \ref{fig:qedPbrem} for the
     bremsstrahlung probability dependence on $Q_\perp$.
  }
\end {center}
\end {figure}

\begin {figure}
\begin {center}
  \includegraphics[scale=0.6]{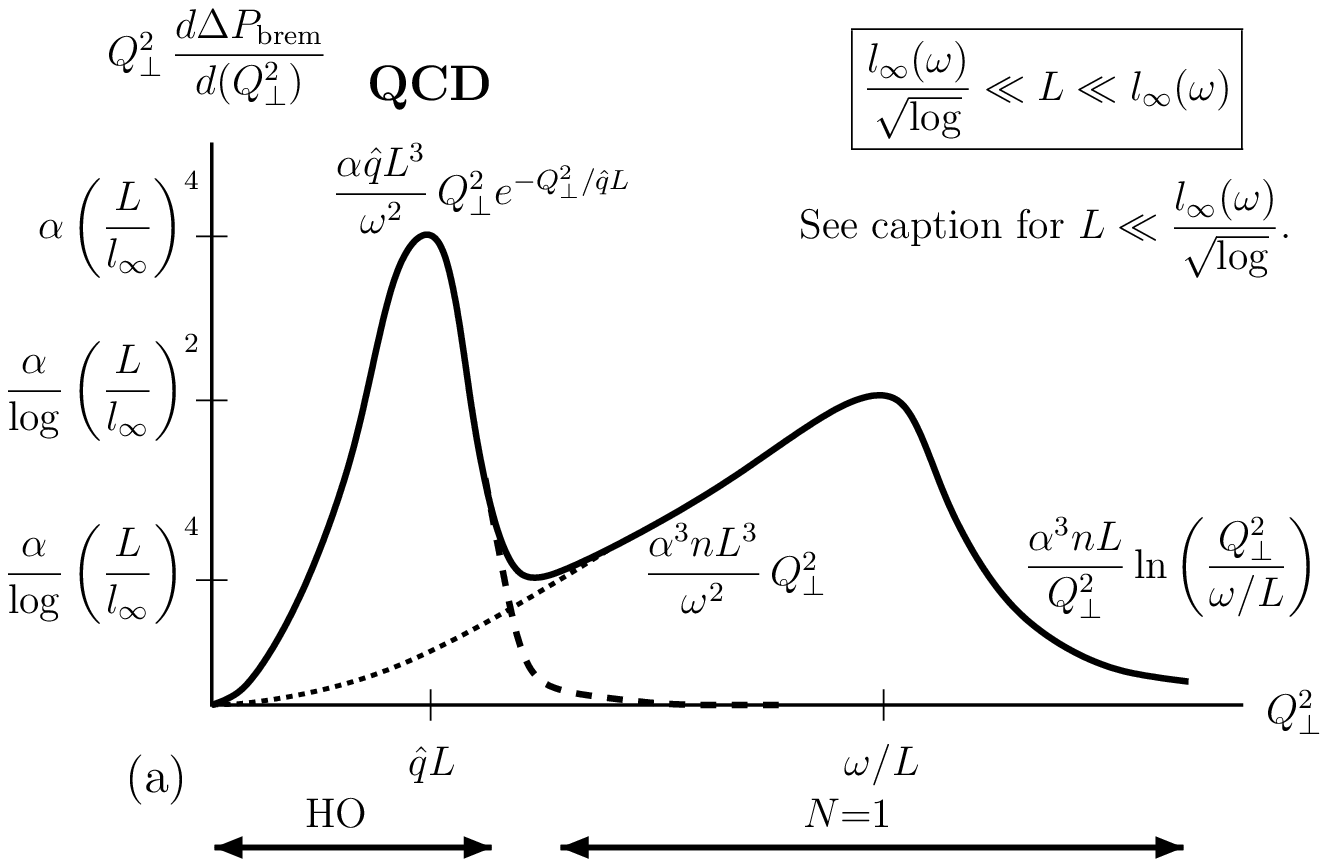}
  \hspace{0.03in}
  \includegraphics[scale=0.6]{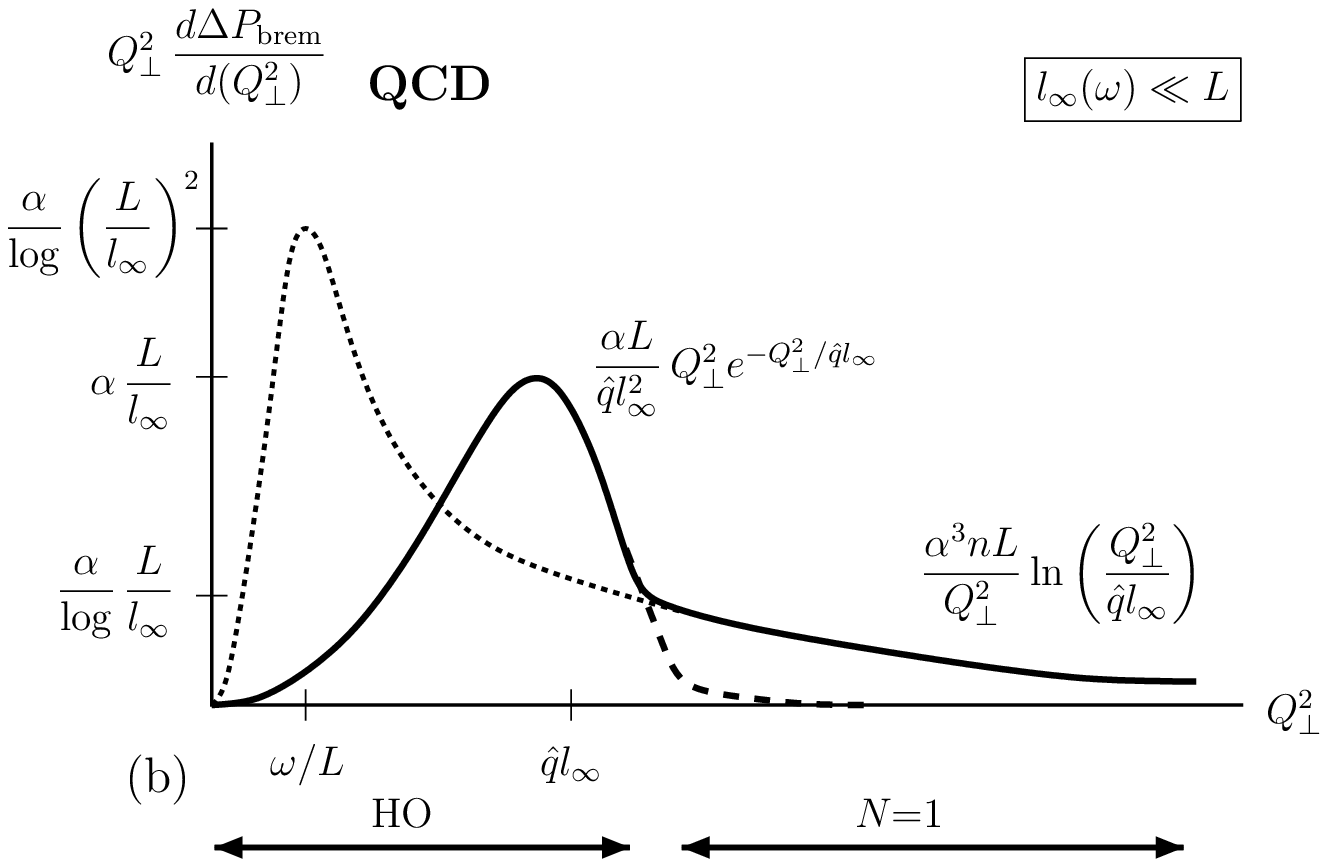}
  \caption{
     \label {fig:qcdPbremQ}
     The QCD analog of Fig.\ \ref{fig:qedPbremQ}.
     For $L \ll \linf(\omega)/\sqrt{\log}$, the right-hand peak
     in (a) is the higher one.
  }
\end {center}
\end {figure}

The analysis of the $Q_\perp$ dependence of the bremsstrahlung
problem is basically the same as in QED, but with the modifications
described above concerning the formation length.  The QCD
versions of Figs.\ \ref{fig:qedLPM}, \ref{fig:qedPbrem} and
\ref{fig:qedPbremQ} are given by \ref{fig:qcdLPM}--\ref{fig:qcdPbremQ}.
The only change in these figures is the replacement of $E$ by
$\omega$ and the clarification that $\hat q$ is $\qhatA$ in the case
$\omega \ll E$.  There are group factors associated
with each power of $\alpha$, but I will not keep track of these
in the figures.  (Similarly, I will not distinguish between
${\cal N}$ and the density $n$ in figures.)
The nature of the collinear logarithm in the QCD case
is reviewed in Appendix \ref{app:QCDcollinear}.

A quick check can be made of the parametric estimate
$\epsilon \sim \qhatA^2 L^4/\omega^2$ shown in Fig.\ \ref{fig:qcdLPM}a
for typical
scatterings $Q_\perp^2 \sim (Q_\perp^2)_{\rm typ}$ for thin media.
The medium effect on the bremsstrahlung probability is then of order
\begin {equation}
   C_s \alpha \epsilon \sim \frac{C_s \alpha \qhatA^2}{\omega^2} \, L^4
\label {eq:epscheck}
\end {equation}
by (\ref{eq:epsdef}), where here I've included the factor of $C_s$
associated with the $\alpha$ for the coupling of the bremsstrahlung
gluon.  Typical scatterings are described by the HO approximation, and
(\ref{eq:epscheck}) correctly reproduces the parametric dependence of
the known HO result (\ref{eq:spectrumHO}) for the thin media modification to
the spectrum of gluon bremsstrahlung.

Now return to the general problem.
Evaluating the integration over $Q_\perp$ by the
peak heights in Fig.\ \ref{fig:qcdPbremQ}, the dependence of
the medium modification $\Delta P_{\rm brem}$ on medium thickness
$L$, for bremsstrahlung gluons with a frequency of order $\sim \omega$,
is given in Fig.\ \ref{fig:qcdPbremL}, but this time with the QCD
value (\ref{eq:linfQCD2}) for $\linf(\omega)$ instead of the QED
version.

\begin {figure}
\begin {center}
  \includegraphics[scale=0.6]{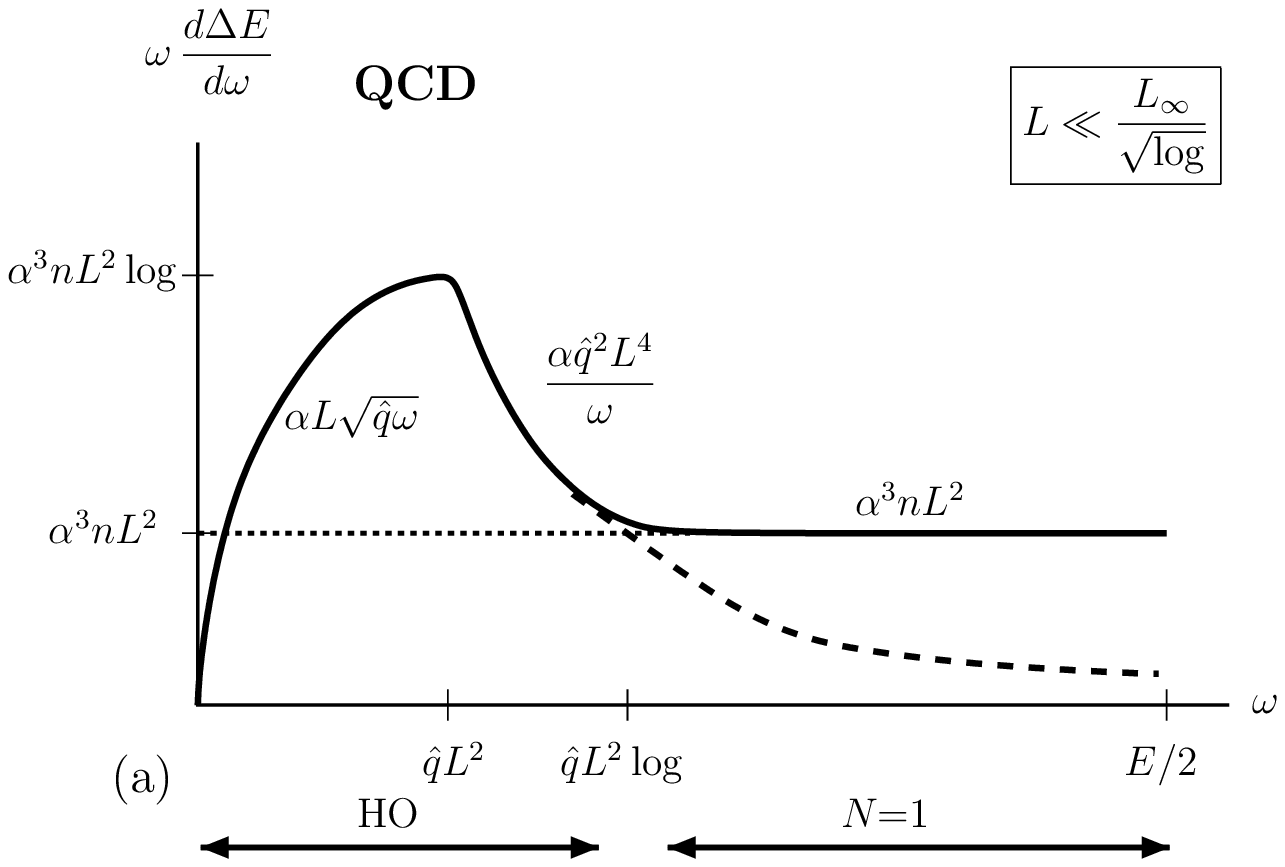}
  \hspace{0.1in}
  \includegraphics[scale=0.6]{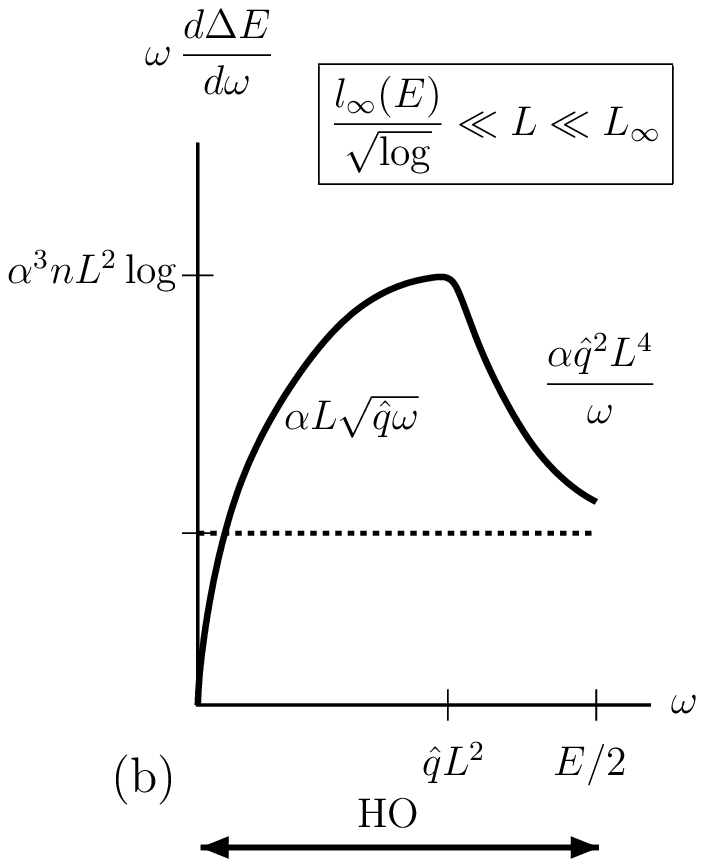}
  \hspace{0.1in}
  \includegraphics[scale=0.6]{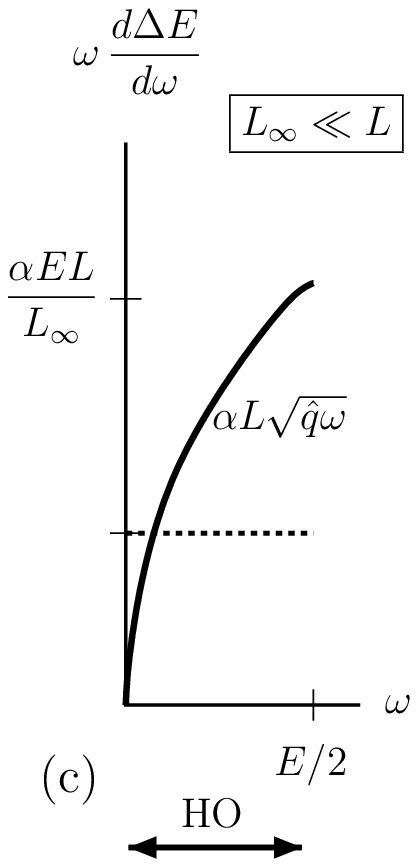}
  \caption{
    \label {fig:qcdE}
    Medium modification to QCD energy loss $\Delta E$:
           contribution of bremsstrahlung gluons with
           frequency $\sim \omega$.
    This is the QCD analog of Fig.\ \ref{fig:qedE}.
    }
\end {center}
\end {figure}

From the peak heights of Fig.\ \ref{fig:qcdPbremQ} or equivalently from
the results for $\Delta P_{\rm brem}(\omega)$ in Fig.\ \ref{fig:qcdPbremL},
one may extract Fig.\ \ref{fig:qcdE} showing the
medium effect on the energy loss spectrum as a function of
frequency $\omega$.
This figure is qualitatively very different from the QED version
of Fig.\ \ref{fig:qedE} because of the qualitative difference in
$\linf(\omega)$.  In QED, small $\omega$ leads to larger formation
lengths and so more LPM suppression, which is why the QED figure
was dominated by $\omega$ of order the largest scale, $\omega \sim E$.
In QCD, small $\omega$ leads to smaller formation lengths and so
less LPM suppression, which is why for thin media the QCD figure
has significant contributions from $\omega \ll E$.

Figs.\ \ref{fig:qcdE}b--c show that the typical scattering processes
captured by the HO approximation dominate (at leading-log order) not
only for length $L$ large compared to the typical infinite-medium
formation length $L_\infty$, but also in the entire range $L \gg
L_\infty/\sqrt{\log}$, which includes $L \sim L_\infty$.
For the case $L \ll L_\infty/\sqrt{\log}$ of Fig.\ \ref{fig:qcdE}a, the
situation is more complicated.  Obtaining the total $\Delta E$
corresponds to integrating the curve in Fig.\ \ref{fig:qcdE} with
$d(\ln\omega) = d\omega/\omega$.  The peak in
Fig.\ \ref{fig:qcdE}a gives an HO-dominated contribution of
order the peak height,
\begin {equation}
   C_s \alpha \qhatA L^2
   \sim
   C_s \ca \alpha^3 {\cal N} L^2 \ln\left( \frac{\qhatA L}{\md^2} \right)
   \qquad
   \mbox{(HO contribution).}
\label {eq:HOlog}
\end {equation}
The long flat plateau at larger $\omega$ gives a contribution of
order the height of the plateau times a logarithm of its range:
\begin {equation}
   C_s \ca \alpha^3 {\cal N} L^2 \times
     \int_{\sim \qhatA L^2 \log}^{\sim E} \frac{d\omega}{\omega}
   \sim C_s \ca \alpha^3 {\cal N} L^2 \ln\left( \frac{E}{\qhatA L^2} \right) .
\label {eq:N1log}
\end {equation}
Over the frequency range
$\qhatA L^2 \log \ll \omega \ll E$
of this plateau, the
value of $\omega \> d(\Delta E)/d\omega$ is well approximated
by the $N{=}1$ approximation, with the scattering events which
dominate the medium effect having the form of Fig.\ \ref{fig:scatt}b.
Adding the two contributions (\ref{eq:HOlog}) and (\ref{eq:N1log})
together gives the final result
(\ref{eq:dE}) quoted in the introduction.  As described there, the
HO contribution continues to dominate the $N{=}1$ contribution
for $L$ all the way down until the
$L_* \sim (\md^4/\qhatA E)^{1/6} L_\infty$ of (\ref{eq:Lstar}).
This is parametrically very different from the QED case
of Fig.\ \ref{fig:qedEL}, where
HO processes dominate $\Delta E$
only down until $L \sim L_\infty/\sqrt{\log}$.

It is important to note that there is no ($N{=}1$)-like
contribution for $\omega \ll \qhatA L^2$ in
Fig.\ \ref{fig:qcdEL}a that is separate from the HO
contribution and that would parametrically reduce the lower scale
$\qhatA L^2$ in the
logarithm of (\ref{eq:N1log}) to make the sum (\ref{eq:dE}) of
(\ref{eq:HOlog}) and (\ref{eq:N1log}) larger.
To check, I will estimate the contribution $\delta(\Delta E)$
to $\Delta E$ from ($N{=}1$)-like scattering (Fig.\ \ref{fig:scatt}b)
with $\omega \lesssim \qhatA L^2$, as opposed to HO-like scattering
(Fig.\ \ref{fig:scatt}a).
The range $\omega \ll \qhatA L^2$
in Fig.\ \ref{fig:qcdE} corresponds to $L \gg \linf(\omega)$
and so to Fig.\ \ref{fig:qcdPbremQ}b.
The contribution to $\Delta P_{\rm brem}$ from the $N{=}1$ part of
that curve is of order the height in Fig.\ \ref{fig:qcdPbremQ}b where
the HO and $N{=}1$ curves meet:
\begin {equation}
   \delta(\Delta P_{\rm brem})
   \sim \frac{C_s \alpha}{\ln(\qhatA L/\md^2)} \, \frac{L}{\linf(\omega)}
   \sim \frac{C_s \alpha L}{\ln(\qhatA L/\md^2)}
       \sqrt{ \frac{\qhatA}{\omega} } \,.
\end {equation}
Multiplying by $\omega$ and integrating over the range
$\omega \lesssim \qhatA L^2$ under discussion gives an
additional $N{=}1$ contribution to $\Delta E$ of
\begin {equation}
   \delta(\Delta E)
   \sim \Bigl[ \omega \, \delta(\Delta P_{\rm brem})
        \Bigr]_{\omega\sim\qhatA L^2}
   \sim \frac{C_s \alpha \qhatA L^2}{\ln(\qhatA L/\md^2)}
   \sim C_s \ca \alpha^3 {\cal N} L^2 .
\end {equation}
This contribution from $(N{=}1)$-like events with
$\omega \lesssim \qhatA L^2$ is sub-leading in logarithms
compared to the HO contribution (\ref{eq:HOlog}) and
the total result (\ref{eq:dE}), and so it can be ignored in
a leading-log analysis.


\section{Comparison to Zakharov's Analysis}
\label {sec:Zakharov}

\subsection {The Puzzle}

In Ref.\ \cite{ZakharovResolution}, Zakharov argued that the HO approximation
should be expected to break down when $L$ is less than or order the
infinite-volume formation time.  In this paper, I have argued that
the HO approximation does a little better than that if one
consistently treats logarithms as large.
For the case of the medium modification to the
bremsstrahlung spectrum, depicted in Fig.\ \ref{fig:qcdPbremQ},
the HO approximation dominates as long as
$L \gg \linf(\omega)/\sqrt{\log}$, which includes
$L \sim \linf(\omega)$.
In this section, I will paraphrase Zakharov's argument and
resolve the slight difference in conclusion.

First, I need to briefly review the formalism for doing a full
calculation of the gluon bremsstrahlung spectrum \cite{BSZ}, which was
originally developed for finite media
by Baier, Dokshitzer, Mueller, Peigne, and Schiff
\cite{BDMPS1,BDMPS2,BDMPS3,BDMS} and by Zakharov
\cite{Zakharov1,Zakharov2}.
A brief summary in my own notation, which is suited for
discussing problems where the particles in the medium are not
fixed scatterers, can be found in Ref.\ \cite{timelpm1}.
The spectrum is given by \cite{timelpm1}
\begin {multline}
   \omega \, \frac{d}{d\omega}(I-I_{\rm vac}) =
   \frac{\alpha x \, P_{s{\to}{\rm g}}(x)}{[x(1-x) E]^2} \,
   \Real
   \int_0^\infty dt_1 \int_{t_1}^\infty dt_2 \>
\\
   \Bigl[
     \grad_{\B_1} \cdot \grad_{\B_2}
     \bigl\{
       G(\B_2,t_2;\B_1,t_1)
       -
       G_{\rm vac}(\B_2,t_2;\B_1,t_1)
     \bigr\}
   \Bigr]_{B_1=B_2=0} ,
\label {eq:general}
\end {multline}
where $G(\B_2,t_2;\B_1,t_1)$ is the Green's function for a
two-dimensional quantum mechanics problem with the
time-dependent, non-hermitian Hamiltonian%
\footnote{
  Zakharov uses the letter $\bm\rho$ for what I call $\B$.
  For a complete translation table, see the appendix of
  Ref.\ \cite{timelpm1}.
}
\begin {equation}
   H(t) = \delta E(\p_B) - i \Gamma_3(\B,t) .
\label {eq:H0}
\end {equation}
Here $\delta E$ describes the energy difference
$(E_{s,\p} + E_{{\rm g},\k}) - E_{s,\p+\k}$ between
(i) a high-energy parton of momentum ${\bm P} = \p+\k$ and
energy $E = P$ and (ii)
the same parton with momentum $\p$ plus a bremsstrahlung gluon with
momentum $\k$.  In the high-energy limit, if we ignore masses,
it can be written as
\begin {equation}
  \delta E(\p_B)
  \simeq
    \frac{p_B^2}{2 x (1-x) P}
  \equiv \frac{p_B^2}{2 M}
\end {equation}
where $\p_B \equiv (p \k_\perp - k\p_\perp)/P$ is the transverse
momentum conjugate to the $\B$ of (\ref{eq:H0})
and $M \equiv x(1-x)P$ is the ``mass'' of the two-dimensional
Schr\"odinger problem:
\begin {equation}
   H(t) \simeq \frac{p_B^2}{2 M} - i \Gamma_3(\B,t) .
\label {eq:H}
\end {equation}
For fixed
$x$, this $\p_B$ is proportional to the angle between $\k$ and $\p$.
The second term in (\ref{eq:H}) is
\begin {equation}
  \Gamma_3(\B,t) =
  \half \ca \, \bar\Gamma_2(\B,t)
  + (C_s - \half\ca) \, \bar\Gamma_2(x\B,t)
  + \half\ca \, \bar\Gamma_2\bigl((1-x)\B,t\bigr) ,
\label {eq:Gamma3}
\end {equation}
where
\begin {equation}
   \bar\Gamma_2(\b,t) \equiv
   \int d^2 q_\perp \>
   \frac{d\bar\Gamma_{\rm el}(t)}{d^2q_\perp}
   \,
   (1 - e^{i \b \cdot \q_\perp})
   .
\label {eq:Gamma2}
\end {equation}
Here $\bar\Gamma_{\rm el}$ is defined by
$\Gamma_{\rm el} \equiv C_R \bar\Gamma_{\rm el}$.
That is, it is the elastic scattering rate without the group
factor $C_R$ associated with the particle being scattered.

The high-energy limit corresponds to large mass $M$ in the
two-dimensional Hamiltonian (\ref{eq:H}) and so will be
determined by the small $B$ behavior of the ``potential''
$-i \Gamma_3(\B,t)$.  Naively, (\ref{eq:Gamma2}) for small $b$
gives
\begin {equation}
   \bar\Gamma_2(\b,t) \simeq \qhat b^2,
\label {eq:HOapprox}
\end {equation}
where $C_R \qhat$ is formally the {\it average}\/ momentum transfer
per unit length rather than the {\it typical} transfer used
throughout this paper.
The actual small $b$ behavior of $\Gamma_2$ is proportional to
$b^2 \ln(\md^2 b^2)$, not $b^2$, which is reflected by the UV divergence of
the integral (\ref{eq:Qperp}) for average transverse momentum transfer.
Cutting off this divergence by replacing $C_R \qhat$ by the
{\it typical}\/ momentum transfer per unit length, as in (\ref{eq:qhat}),
corresponds to the harmonic oscillator approximation, so named because
of the form of (\ref{eq:HOapprox}).

In contrast, another analytic approach to solving the problem is
to keep the full original form of (\ref{eq:H}) and instead do
perturbation theory in powers of the
(imaginary-valued) potential $-i \Gamma_3$.
This is the formal version of the opacity expansion.

Alternatively, {\it both}\ approximations can be made.
Consider the case of the brick problem.
If one first makes the HO approximation (\ref{eq:HOapprox}) and
then makes the opacity expansion, the opacity expansion is simply
the expansion of the HO result (\ref{eq:IHO}) in powers of
$|\omega_0^2| \propto \qhat \propto -i \Gamma_3$:
\begin {equation}
   \omega \, \frac{d}{d\omega}(I-I_{\rm vac})_{\rm HO} =
   \frac{\alpha}{\pi} \, x \, P_{s{\to}{\rm g}}(x) \,
   \left[
     \tfrac{1}{12} |\omega_0|^4 L^4
     - \tfrac{17}{2520} |\omega_0|^8 L^8
     + \cdots 
   \right] .
\label {eq:HOexpand0}
\end {equation}
Parametrically, this expansion has the form
\begin {align}
   \omega \, \frac{d}{d\omega}(I-I_{\rm vac})_{\rm HO}
   &\sim
   C_s \alpha \left[
     \# \left(\frac{\qhatA L^2}{\omega}\right)^2
     + \# \left(\frac{\qhatA L^2}{\omega}\right)^4
     + \cdots
   \right]
\nonumber\\
   &\sim
   C_s \alpha \left[
     \# \left(\frac{L^2}{[\linf(\omega)]^2}\right)^2
     + \# \left(\frac{L^2}{[\linf(\omega)]^2}\right)^4
     + \cdots
   \right]
\nonumber\\
   &\sim
   C_s \alpha \left[
     \# \left(\frac{\ca \alpha^2 {\cal N} L^2 \log}{\omega}\right)^2
     + \# \left(\frac{\ca \alpha^2 {\cal N} L^2 \log}{\omega}\right)^4
     + \cdots
   \right] ,
\label {eq:HOexpand}
\end {align}
with the logarithms defined as in (\ref{eq:log}).
The condition for the perturbative expansion of
(\ref{eq:HOapprox}) to be useful is that successive terms
get smaller and smaller,
which parametrically is the condition that $L \ll \linf(\omega)$.
As noted by Zakharov \cite{ZakharovResolution}, there is
no first-order term (no term proportional to $\omega_0^2$ and
so proportional to the interaction $\Gamma_3$) in the expansion
(\ref{eq:HOexpand0}).  But if $L \ll \linf(\omega)$ so that
a perturbative treatment of the quantum mechanical problem is
valid, then why not forgo the HO approximation and just use
the full, original potential $-i\Gamma_3$.  At first order,
one then obtains the $N{=}1$ result (\ref{eq:spectrumN1}),
so that
\begin {equation}
   \omega \, \frac{d}{d\omega}(I-I_{\rm vac}) \sim
   C_s \alpha \left[
     \# \, \frac{\ca \alpha^2 {\cal N} L^2}{\omega}
     + \cdots
   \right] .
\end {equation}
The fact that a perturbative expansion of the HO result should work whenever
$L \ll \linf(\omega)$, yet the HO approximation is clearly missing the
first-order term in this limit, makes it seem like the HO
approximation must be untrustworthy whenever $L \ll \linf(\omega)$.


\subsection {Reconciliation}

The absence of the first-order term in the expansion of the HO
result can be illuminated if one separates out from
Eq.\ (\ref{eq:general}) the step of taking the real part.
The origin of the HO result (\ref{eq:IHO}) is actually
\begin {equation}
   \omega \, \frac{d}{d\omega}(I-I_{\rm vac})_{\rm HO} =
   \frac{\alpha}{\pi} \, x \, P_{s{\to}{\rm g}}(x) \,
   \Re \left[ \ln\cos(\omega_0 L) \right] .
\end {equation}
Correspondingly, using $\omega_0^2 = -i |\omega_0|^2$,
the perturbative expansion is
\begin {equation}
   \omega \, \frac{d}{d\omega}(I-I_{\rm vac})_{\rm HO} =
   \frac{\alpha}{\pi} \, x \, P_{s{\to}{\rm g}}(x) \,
   \Re \left[
     \tfrac{i}2 |\omega_0|^2 L^2
     + \tfrac1{12} |\omega_0|^4 L^4
     + \tfrac{i}{45} |\omega_0|^6 L^6
     - \tfrac{17}{2520} |\omega_0|^8 L^8
     + \cdots
   \right] ,
\end {equation}
which shows the missing odd terms in the expansion.
This corresponds to
\begin {equation}
   \omega \, \frac{d}{d\omega}(I-I_{\rm vac})_{\rm HO}
   \sim
   C_s \alpha \Re \left[
     i \, \# \frac{\ca \alpha^2 {\cal N} L^2}{\omega}
           \ln \left( \frac{\qhatA L}{\md^2} \right)
     + \# \left(\frac{\ca \alpha^2 {\cal N} L^2 \log}{\omega}\right)^2
     + \cdots
   \right]
\label {eq:HOshow}
\end {equation}
In contrast, the full $N{=}1$ perturbative calculation turns out
to give
\begin {align}
   \omega \, \frac{d}{d\omega}(I-I_{\rm vac})_{N{=}1}
   &\simeq
   \frac{\ca \alpha^3 {\cal N}}{\omega} \, L^2 \,
   x  \, P_{s{\to}{\rm g}}(x) \,
   \Re \left[ i
     \ln \left( \frac{\md^2 L}{\# i M} \right)
   \right]
\nonumber\\
   &=
   \frac{\ca \alpha^3 {\cal N}}{\omega} \, L^2 \,
   x  \, P_{s{\to}{\rm g}}(x) \,
   \Re \left[ i
     \ln \left( \frac{\md^2 L}{\# M} \right)
     + \frac{\pi}{2} 
   \right]
\end {align}
in the limit of large logarithms.
Taking the real part and the small $x$ limit, this reproduces
(\ref{eq:spectrumN1}).  Parametrically,
\begin {align}
   \omega \, \frac{d}{d\omega}(I-I_{\rm vac})_{N{=}1}
   &\sim
   C_s \alpha \Re \left[
     i \, \# \frac{\ca \alpha^2 {\cal N} L^2}{\omega}
           \ln \left( \frac{\md^2}{i \omega} \right)
   \right]
\nonumber\\
   &\sim
   C_s \alpha \Re \left[
     i \# \frac{\ca \alpha^2 {\cal N} L^2}{\omega}
     \left\{
        \ln \left( \frac{\md^2}{\omega} \right) - i \#
     \right\}
   \right]
\label {eq:N1show}
\end {align}

Comparing (\ref{eq:HOshow}) and (\ref{eq:N1show}) {\it before}\/ taking the
real part, note that the first-order terms are parametrically the
same except that the arguments of the large logarithms are different.
When the real part is taken, however, nothing survives of the
first term in the expansion (\ref{eq:HOshow}) of the HO result, but a
term sub-leading in large logarithms survives from the $N{=}1$
result.  The moral is that the structure of the $N{=}1$ and
HO results for thin media are not very different before one takes
the real part.  After the real part, a calculation which included
contributions from both HO and $N{=}1$ physics, as described in
this paper, would be expected to produce a result of the form
\begin {equation}
   \omega \, \frac{d}{d\omega}(I-I_{\rm vac})_{\rm HO}
   \sim
   C_s \alpha \left[
     \frac{\ca \alpha^2 n L^2}{\omega}
     + \# \left(\frac{\ca \alpha^2 {\cal N} L^2 \log}{\omega}\right)^2
     + \cdots
   \right]
\label {eq:reconcile}
\end {equation}
This is consistent with the $L \lesssim \linf(\omega)$
behavior of $\Delta P_{\rm brem}$ found in this paper and
shown in Fig.\ \ref{fig:qcdPbremL}.


\subsection {Some Differences}

It is important to note, however, that a perturbative calculation
to second order in $-i\Gamma_3$ would not give precisely
(\ref{eq:reconcile}) with the HO logarithm (\ref{eq:log}),
\begin {equation}
   \log \equiv \ln\left( \frac{\qhatA L}{\md^2} \right)
        \sim \ln\left( \frac{\ca \alpha^2 {\cal N}L}{\md^2} \right) .
\label {eq:log2}
\end {equation}
When discussing the perturbative expansion (\ref{eq:HOexpand}) of
the HO result, I {\it first}\/ made the HO approximation
(\ref{eq:HOapprox}) and treated $\qhat$ as a constant given by
(\ref{eq:qhat}).  Only then did I expand in powers of $-i\Gamma_3$.
If I instead forgo the HO approximation, then the expansion
in $-i\Gamma_3$ (the opacity expansion)
is equivalent to an expansion in powers of the
medium density $\sim {\cal N}$,
if for this purpose I treat the Debye screening
mass $\md$ as a variable independent from ${\cal N}$.
However, (\ref{eq:reconcile}) is not a simple power series expansion
in ${\cal N}$, because there is a factor of ${\cal N}$ inside the argument to
the logarithm (\ref{eq:log2}).  The second-order HO term in
(\ref{eq:reconcile}) can therefore only arise from a resummation
of many terms of the opacity expansion.%
\footnote{
  Readers may wonder how a logarithm of the form $\ln(c{\cal N})$ could
  possibly arise from any power series in ${\cal N}$,
  since $\ln(c{\cal N})$
  is not expandable as a power series.  Keep in mind that the
  form (\ref{eq:log2}) of the logarithm is only meant to be valid
  in the limit that the argument of the logarithm is large.
  So, as an example, $\ln(1+c{\cal N}) \simeq \ln(c{\cal N})$
  when the argument is
  large, but $\ln(1+c{\cal N})$ has a series expansion in ${\cal N}$.
}

In this paper, I will not attempt to explore in detail how physics
associated with the HO approximation can be seen to arise from
resummation of terms in the opacity expansion.
But I hope that the discussion of this section gives some
insight into how the earlier results of this paper, based on
more physical arguments, can be consistent formally with the small $L$
expansions of the $N{=}1$ and HO approximations.


\begin{acknowledgments}

I am indebted to B.G. Zakharov, Guy Moore, Al Mueller, and J.P. Nolan
for useful discussions.
This work was supported, in part, by the U.S. Department
of Energy under Grant No.~DE-FG02-97ER41027.

\end{acknowledgments}

\appendix

\section {The form of \boldmath$dP_{\rm scatt}/d(Q_\perp^2)$}
\label{app:Pscatt}

The qualitative form of Fig.\ \ref{fig:Pscatt} --- a diffusion
peak from typical scatterings with a single-scattering tail due to
rare scatterings --- has a very long history.
For example, a simplified version of Molliere's theory of multiple scattering
was given by Bethe in 1953 \cite{Bethe}.%
\footnote{
  For other references, see Sec.\ 27.3 of the 2008 Review of
  Particle Physics \cite{pdb08}.
}
In the current context of leading-log approximations to
jet $Q_\perp$ broadening in quark-gluon
plasmas, it has been addressed
previously in Sec.\ 3.1 of Ref.\ \cite{BDMPS3} and is
nicely reviewed in Appendix A of Ref.\ \cite{PeigneSmilga} for
the particular model of scattering where $d\Gamma_{\rm el}/d(q_\perp^2)$ is
taken to be proportional to $(q_\perp^2 + \md^2)^{-2}$.
For the sake of completeness, I will
review here the important elements for the current work
in a model-independent way.
Specifically, the important points for my argument in this
paper are that (i) $dP_{\rm scatt}/d(Q_\perp^2) \sim 1/\hat q L$ for
$Q_\perp^2 \ll \hat q L$ (the peak height in Fig.\ \ref{fig:Pscatt}),
and (ii) $dP_{\rm scatt}/d(Q_\perp^2) \sim \alpha^2 n L/Q_\perp^4$
for $Q_\perp^2 \gg \hat q L$ (the form of the large $Q_\perp$ tail).

The transition between these two behaviors occurring for
$\hat q L \lesssim Q_\perp^2 \lesssim \hat q L \ln(\log)$ in
Fig.\ \ref{fig:Pscatt} is interesting but unimportant to
my conclusions.
In this paper, I treat logarithms as large but I treat logarithms
of logarithms $\ln(\log)$ as $O(1)$.  So $Q_\perp^2 \gg \hat q L$
refers to the tail of Fig.\ \ref{fig:Pscatt} and not to any part of
the transition region.

Though not needed for the present work, I will also provide reference
to a rigorous mathematical generalization of the
central limit theorem which demonstrates that the soft-scattering peak
indeed approaches a Gaussian form in the limit of a large number
of collisions.


\subsection {Review of general multiple scattering formula}

Let
\begin {equation}
   f(Q_\perp,t) \equiv \frac{dP_{\rm scatt}}{d^2Q_\perp}
   = \frac1{\pi} \frac{dP_{\rm scatt}}{d(Q_\perp^2)}
\end {equation}
be the two-dimensional probability distribution of
$\Q_\perp$ at time $t$.  I follow the standard development of
multiple scattering by writing the evolution equation for $f$,
which is
\begin {equation}
  \partial_t f(\Q_\perp,t)
  = \int_{\q_\perp} \rho(\q_\perp) [f(\Q_\perp{-}\q_\perp,t) - f(\Q_\perp,t)]
\label {eq:evolvef}
\end {equation}
in a uniform medium, where
\begin {equation}
  \rho(q_\perp) \equiv 
  \frac{d\Gamma_{\rm el}(\q_\perp)}{d^2 q_\perp}
\end {equation}
is the two-dimensional probability density for acquiring a transverse
momentum kick of $\q_\perp$ in a single, individual collision.
The first term on the right-hand side of (\ref{eq:evolvef}) is
a gain term, representing momentum change from $\Q_\perp{-}\q_\perp$
to $\Q_\perp$.  The second term is a loss term, representing
change from $\Q_\perp$ to $\Q_\perp{+}\q_\perp$.
Take the initial condition $f(\Q_\perp,0) = \delta^{(2)}(\Q_\perp)$.
The equation is solved
by Fourier transformating to
\begin {equation}
  \partial_t \tilde f(\b,t)
  = [\tilde\rho(\b) - \tilde\rho(\bm 0)] f(\b,t)
\end {equation}
with initial condition $\tilde f(\b,0) = 1$ and solution
\begin {equation}
  \tilde f(\b,t) = e^{-[\tilde\rho(\bm0) - \tilde\rho(\b)]t} .
\end {equation}
Fourier transforming back,
\begin {equation}
  \frac{dP_{\rm scatt}}{d^2Q_\perp}
    = \int_{\b} e^{-[\tilde\rho(\bm0) - \tilde\rho(\b)]t}
    e^{-i \b\cdot\Q_\perp} ,
\label {eq:Pmultiple}
\end {equation}
where
\begin {equation}
  \tilde\rho(\bm 0) - \tilde\rho(\b) =
  \int_{\q_\perp} 
  \frac{d\Gamma_{\rm el}(\q_\perp)}{d^2 q_\perp}
  (1 - e^{i \b\cdot\q_\perp}) .
\end {equation}


\subsection {Examples}

As an example, consider the weak-coupling result \cite{AGK}%
\footnote{
  For a brief overview
  in the notation used here, see Sec. II A of Ref.\ \cite{ArnoldXiao}.
}
\begin {equation}
  \frac{d\Gamma_{\rm el}}{d^2 q_\perp}
  \simeq \frac{C_R g^2 T \md^2}{(2\pi)^2 q_\perp^2(q_\perp^2+\md^2)}
\label {eq:ex1}
\end {equation}
which holds for $q_\perp \ll T$.  The result is not significantly
different for $q_\perp \gg T$ \cite{ArnoldDogan,ArnoldXiao},
so I shall take it as an example
for discussing the entire range of $q_\perp$.
The $1/q_\perp^2$ behavior for $q_\perp \ll \md$ is due to
magnetic scattering, which is not completely screened by the Debye
effect.
The Fourier transform of (\ref{eq:ex1}) gives \cite{AMYimpact}
\begin {equation}
   \tilde\rho(\bm0) - \tilde\rho(\bm b)
   = \frac{C_R g^2 T}{2\pi}
      \left[ K_0(\md b) - \ln\left(\frac{2}{\md b}\right) + \gammaE \right]
   .
\end {equation}
Note that for $\md b \ll 1$ this becomes
\begin {equation}
   \tilde\rho(\bm0) - \tilde\rho(\b)
   \simeq \frac{C_R g^2 T}{8\pi}
     (\md b)^2
     \left[ \ln\left(\frac{2}{\md b}\right) - \gammaE + 1 \right]
   ,
\label {eq:ex1small}
\end {equation}
and for $\md b \gg 1$ it is
\begin {equation}
   \tilde\rho(\bm0) - \tilde\rho(\b)
   \simeq
   \frac{C_R g^2 T}{2\pi}
     \left[ - \ln\left(\frac{2}{\md b}\right) + \gammaE \right]
   \to +\infty
   \quad \mbox{as $b \to \infty$.}
\end {equation}

Alternatively, consider a popular model used in this subject, which is
\begin {equation}
  \frac{d\Gamma_{\rm el}}{d^2 q_\perp}
  \to \frac{C_R g^4 {\cal N}}{(2\pi)^2 (q_\perp^2+\md^2)^2} ,
\label {eq:ex2}
\end {equation}
where ${\cal N}$ is the number density weighted by appropriate group factors.%
\footnote{
  In detail, I am using the notation of Ref.\ \cite{ArnoldXiao}.
}
Fourier transforming, one finds
\begin {equation}
  \tilde\rho(\b) =
  \frac{C_R g^4 {\cal N}}{4\pi\md} \, b \, K_1(\md b) ,
\end {equation}
so that
\begin {equation}
  \tilde\rho(\bm 0) - \tilde\rho(\b) =
  \frac{C_R g^4 {\cal N}}{4\pi\md^2} \,
     \left[ 1 - \md b \, K_1(\md b) \right] .
\end {equation}
For $\md b \ll 1$,
\begin {equation}
   \tilde\rho(\bm0) - \tilde\rho(\b)
   \simeq \frac{C_R g^4 {\cal N}}{8\pi}
     \, b^2
     \left[ \ln\left(\frac{2}{\md b}\right) - \gammaE + \frac12 \right]
   ,
\label {eq:ex2small}
\end {equation}
and for $\md b \gg 1$ it is
\begin {equation}
   \tilde\rho(\bm0) - \tilde\rho(\b)
   \simeq
   \tilde\rho(\bm0)
   = \frac{C_R g^4 {\cal N}}{4\pi\md^2}
   \qquad \mbox{as $b \to \infty$.}
\label {eq:binfty2}
\end {equation}

The $b^2 \ln(\md b)$ at small $b$ in both examples is a universal
result of having a $d\Gamma_{\rm el}/d^2q_\perp$ that falls as
$1/q_\perp^4$ at large $q_\perp$.  This in turn is a universal
feature of point-particle scattering that is Coulomb at short distances.
Note that the two formulas (\ref{eq:ex1small}) and (\ref{eq:ex2small})
have the same size parametrically, so it does not matter which we
use if we are interested in a parametric analysis.%
\footnote{
   Eq.\ \ref{eq:ex2small} is the correct formula at high enough energy
   that typical individual scatterings have $q_\perp \gg T$.
   See, for example, the discussion in Ref.\ \cite{ArnoldXiao}.
}
In general,
\begin {equation}
   \tilde\rho(\bm0) - \tilde\rho(\b)
   \sim C_R g^4 {\cal N}
     b^2
     \ln\left(\frac{1}{\md b}\right)
   \quad
   \mbox{for $\md b \ll 1$.}
\label {eq:exsmall}
\end {equation}

In the second example,
the finite $b{\to}\infty$ limit (\ref{eq:binfty2})
means that the final Fourier
transform (\ref{eq:Pmultiple}) used to obtain $dP_{\rm scatt}/d^2Q_\perp$
contains a $\delta$-function singularity.  This can be isolated by rewriting
(\ref{eq:Pmultiple}) as
\begin {equation}
  \frac{dP_{\rm scatt}}{d^2Q_\perp}
    = e^{-\tilde\rho(\bm0) \, t} \delta^{(2)}(\b)
      + \int_{\b} \left[ e^{-[\tilde\rho(\bm0) - \tilde\rho(\b)]t}
                         - e^{-\tilde\rho(\bm0) \, t}
                  \right]
    e^{-i \b\cdot\Q_\perp} .
\label {eq:dPreg}
\end {equation}
The coefficient of the $\delta$-function is just the probability
$\exp(-\Gamma_{\rm el} t)$ that there are no scatterings whatsoever.
For large $\Gamma_{\rm el} t$ this can be ignored.
This
$\delta$-function does not appear in the first example
(\ref{eq:ex1}) since in that case
$\Gamma_{\rm el} = \int_{\q_\perp} d^2 \Gamma_{\rm el}/d^2 q_\perp=\infty$
because of the $1/q_\perp^2$ infrared behavior of
(\ref{eq:ex1}) due to magnetic scattering.%
\footnote{
  This infrared divergence will be cut off at $q_\perp \sim g^2 T$ by
  non-perturbative physics.
  The resulting $\delta$-function term
  will in any case be exponentially small when $t \gg 1/\md$,
  which has been assumed throughout this paper.
}
Either way, I will ignore the $\delta$ function in the rest of this
discussion.


\subsection {The height of the peak}

For the height of the peak if Fig.\ \ref{fig:Pscatt}, we just need
to evaluate the regular (i.e.\ non-$\delta$-function) term
in (\ref{eq:dPreg}) at
$Q_\perp{=}0$:
\begin {equation}
  \left(\frac{dP_{\rm scatt}}{d^2Q_\perp}\right)^{\rm reg}_{Q_\perp=0}
  =
    \int_{\b} \left[ e^{-[\tilde\rho(\bm0) - \tilde\rho(\b)]t}
                         - e^{-\tilde\rho(\bm0) \, t}
                  \right] .
\label {eq:dPreg0}
\end {equation}
For large enough $t$, this integral is dominated by small $b$ determined by
$[\tilde\rho(\bm0)-\tilde\rho(\b)] t \sim 1$.
Using (\ref{eq:exsmall}), this is
\begin {equation}
   C_R g^4 {\cal N} t b^2 \ln\left( \frac{1}{\md b}\right) \sim 1,
\end {equation}
and so
\begin {equation}
   b^2
   \sim \left[ C_R g^4 {\cal N} t
                   \ln\left( \frac{C_R g^4 {\cal N} t}{\md^2} \right)
        \right]^{-1}
   \sim \frac{1}{\hat q_R t} .
\label{eq:b2estimate}
\end {equation}
The condition $\md b \ll 1$ that I have used is then satisfied provided
$t \gg \md^2/(C_R g^4 {\cal N})$, which, neglecting group factors, is
of order $1/g^2 T$.  This is just the condition that there are many
$q_\perp{\sim}\md$ scatterings, which I have assumed throughout this paper.
From (\ref{eq:b2estimate}), the size of the integral (\ref{eq:dPreg0})
is then
\begin {equation}
  \left(\frac{dP_{\rm scatt}}{d^2Q_\perp}\right)^{\rm reg}_{Q_\perp=0}
  \sim
  b^2
  \sim \frac{1}{\hat q_R t} .
\end {equation}
Replacing $t$ by $L$, this is just the peak height depicted in
Fig.\ \ref{fig:Pscatt}.

The $Q_\perp{=}0$ result will be a good approximation whenever the
$\exp(i\b\cdot\Q_\perp)$ factor in (\ref{eq:dPreg}) is approximately 1
up to and including $b$ values of order (\ref{eq:b2estimate}).
So $dP_{\rm scatt}/d^2 Q_\perp \sim 1/\hat q_R L$ for
$Q_\perp \ll b^{-1} \sim \sqrt{\hat q_R L}$, just as shown in
Fig.\ \ref{fig:Pscatt}.  For larger $Q_\perp$, the oscillating factor
will cause $dP_{\rm scatt}/d^2 Q_\perp$ to fall.
For much larger $Q_\perp$, the oscillating factor
$\exp(i\Q_\perp\cdot b_\perp)$
causes the integral in (\ref{eq:dPreg}) to be dominated by even
smaller $b$, in which case one may approximate
\begin {equation}
  e^{-[\tilde\rho(\bm0) - \tilde\rho(\b)]t}
  \simeq 1 -[\tilde\rho(\bm0) - \tilde\rho(\b)]t .
\end {equation}
Completing the Fourier transform, this just gives that the
large $Q_\perp$ behavior of $d\Gamma/d^2Q_\perp$ is given by
the single-scattering formula $d\Gamma_{\rm el}/d^2 q_\perp$ with
$\q_\perp = \Q_\perp$, producing the tail of Fig.\ \ref{fig:Pscatt}.


\subsection {The Gaussian shape of the peak}

Ref.\ \cite{BDMPS3} discusses the Gaussian shape of the peak by
noting that $\ln(b^2)$ is a slowly varying function and so making
the approximation of replacing this logarithm by a constant in
(\ref{eq:dPreg}).  This is the harmonic oscillator approximation and
gives a Gaussian result for $dP_{\rm scatt}/d^2Q_\perp$.
Ref.\ \cite{BDMPS3} notes that this approximation breaks down for
the tail, which is generated by the non-analyticity of $\ln b$ at
$b{=}0$.
Some readers may wonder, however, if the argument that the shape
approaches Gaussian can be made more rigorous.
If the distribution $d^2 \Gamma_{\rm el}/q^2q_\perp$ of
single-scattering rates had a finite variance
$\langle q_\perp^2 \rangle < \infty$, then
the the approach to a Gaussian shape would be guaranteed by the
central limit theorem.%
\footnote{
   The limit is non-uniform, which means that for finite $t$ there
   will still be non-Gaussian tails.  However, as $t$ increases,
   the region of $Q_\perp$ over which Gaussian is a good approximation
   becomes larger and larger in units of the width of that Gaussian.
}
Finite variance is a sufficient condition
for the central limit theorem, but it is not a necessary condition.
A necessary and sufficient condition may be found in
Theorem 8.1.3 of Ref.\ \cite{Meerschaert}.%
\footnote{
   Ref.\ \cite{Meerschaert} gives a condition for distributions of
   vectors.  The specialization to one dimensional distributions
   has an older history: see Theorem 1 on p.\ 172 of Ref.\
  \cite{Gnedenko} and references therein.
  Also, the theorem is formulated for a sum of a finite number of
  vectors drawn from the distribution $\rho$.  In our case, one
  should think of $\rho$ as the probability distribution for
  picking up transverse momentum $\q_\perp$ in a very small time
  interval $\Delta t$, and the total $\Q_\perp$ is then the sum of
  these transfers, with the limit $\Delta t \to 0$ taken
  at the end of the day.
}
I will simplify their condition to the radially-symmetric case
of interest here.%
\footnote{
  Ref.\ \cite{Meerschaert}, which also applies
  in the absence of radial symmetry, defines
  $F(\bm x) \equiv
  \int_{\q_\perp} (\bm x\cdot\q_\perp)^2 \, \rho(\q_\perp) \,
                  \theta(1-|\bm x\cdot\q_\perp|)
  $.  Their condition is that there exists (i) a function
  $f(t)$ from $\mathbb{R}^+$ to the set of linear transformations on
  $\mathbb{R}^d$ with $f(\lambda t) \, f(t)^{-1} \to \lambda^{-E} I$
  (where $I$ is the identity operator) as
  $t \to \infty$ for some $E>0$ and for all $\lambda > 0$
  and (ii) another function $R(t)$ from $\mathbb{R}^+ \to \mathbb{R}^+$
  with $R(\lambda t) / R(t) \to \lambda^{2E}$,
  such that (iii)
  $$
    \lim_{t\to\infty} \frac{F(f(t)^{-1} \bm\x_t)}{R(t)} = \phi(\bm\x)
  $$
  for some $\phi(\bm\x) > 0$ whenever $\bm\x_t \to \bm\x$ in
  $\mathbb{R}^d-\{\bm0\}$.
  See pp.\ 96, 125--7, and 293 of Ref.\ \cite{Meerschaert}.
  For the radially symmetric case, $F(\bm x)$ is proportional to
  $|\bm x|^2$ times what I have called $\hat q_{\bm\Lambda}$, and
  their $\x$ is my $\bm n/\Lambda$.
  Workable choices for the specific isotropic case of interest here, where
  $\hat q_\Lambda \propto \ln\Lambda$ at large $\Lambda$ and so
  $F(\x) \propto x^2 \ln(1/x)$ at small $x$,
  are $f(t) = t I$ and $R(t) = t^2 \ln t$.
}
Using the notation of this paper, the condition can be written in terms of
the truncated variance
\begin {equation}
  \hat q_\Lambda \equiv
  \int_{\q_\perp} (\q_\perp\cdot\bm n)^2 \, \rho(q_\perp) \,
                  \theta(\Lambda-|\q_\perp\cdot\bm n|)) ,
\end {equation}
where $\theta(z)$ is the step function,
$\q_\perp$
could be a vector of any dimension, $\rho(\q_\perp)$ is a
probability density in that vector space,
$\Lambda$ is a cut-off, and $\bm n$
is a unit vector in any direction.
The necessary and sufficient condition for the central limit theorem
is that
\begin {equation}
   \lim_{\Lambda\to \infty} 
   \frac{\hat q_{\lambda\Lambda}}{\hat q_{\Lambda}}
   = 1
\end {equation}
for all $\lambda > 0$.
This condition applies to the case relevant in this paper, where
$\hat q_{\Lambda} \propto \ln\Lambda$ in the limit of large $\Lambda$.


\section {More details on suppression \boldmath$\epsilon$ in QED}
\label{app:qed}

\subsection {Review: Why \boldmath$\epsilon \sim (K\cdot\Delta x)^2$ in QED}
\label{app:LPMsupp}
\label{app:collinear}

Here I will briefly review the parametric formula (\ref{eq:LPMsupp})
for the suppression factor $\epsilon$.  For simplicity, I will focus
on QED in the soft bremsstrahlung limit, $\omega \ll E$.
In this limit, the charged particle trajectory can be thought of as
a fixed, classical source $J^\mu(t,\x)$ for the electromagnetic field,
and the bremsstrahlung amplitude is proportional to the Fourier
transform $J^\mu(\omega,\k)$.
For further simplicity, I will focus on a comparison of
a single scattering from the medium, as in
Figs.\ \ref{fig:collinear}a--b, with the vacuum case of
Fig.\ \ref{fig:collinear}c.

The Fourier transform $J^\mu(K)$ of the current
for the trajectory shown in Figs.\ \ref{fig:collinear}a-b is
\begin {align}
   J^\mu(K) &\propto
   \frac{V_{\rm i}^\mu}{K \cdot V_{\rm i}} (e^{i K\cdot X_0}-e^{i K\cdot X_1})
   + \frac{V_{\rm f}^\mu}{K \cdot V_{\rm f}} \, e^{i K \cdot X_1}
\nonumber\\
   &= e^{i K\cdot X_1} \left[
     \frac{V_{\rm i}^\mu}{K \cdot V_{\rm i}} (e^{-i K\cdot \Delta X}-1)
   + \frac{V_{\rm f}^\mu}{K \cdot V_{\rm f}}
   \right] ,
\end {align}
where $K = (\omega,\k) = (k,\k)$ is the photon 4-momentum,
and $V \equiv (1,{\bm v})$ with the charged particle
velocity ${\bm v}$ before
($V_{\rm i}$) and after ($V_{\rm f}$) the medium collision.
The medium effect on the bremsstrahlung probability is then
\begin {equation}
  |\varepsilon \cdot J|^2 - |\varepsilon \cdot J|_{\rm vac}^2
  \propto
  \left|
    \frac{\varepsilon\cdot V_{\rm i}}{K \cdot V_{\rm i}}
          (e^{-i K\cdot \Delta X}-1)
     + \frac{\varepsilon\cdot V_{\rm f}}{K \cdot V_{\rm f}}
  \right|^2
  -
  \left|
     \frac{\varepsilon\cdot V_{\rm f}}{K \cdot V_{\rm f}}
  \right|^2 ,
\label {eq:collinear}
\end {equation}
where $\varepsilon^\mu$ is the photon polarization.
The first term in (\ref{eq:collinear})
corresponds to Figs.\ \ref{fig:collinear}a--b
and the second to Fig.\ \ref{fig:collinear}c.
Expanding the first square and summing over polarizations gives
\begin {equation}
  \sum_\varepsilon \left[
    |\varepsilon \cdot J|^2 - |\varepsilon \cdot J|_{\rm vac}^2
  \right]
  \propto
  2 \left[1 - \cos(K\cdot\Delta X)\right]
  \sum_\varepsilon
  \Re \left[
  \frac{\varepsilon^*\cdot V_{\rm i}}{K \cdot V_{\rm i}}
  \left(
    \frac{\varepsilon\cdot V_{\rm i}}{K \cdot V_{\rm i}}
     - \frac{\varepsilon\cdot V_{\rm f}}{K \cdot V_{\rm f}}
  \right)
  \right] .
\label {eq:collinear2}
\end {equation}
For $K\cdot\Delta X \ll 1$, the bremsstrahlung probability
is therefore proportional to $(K\cdot\Delta X)^2$, as in
(\ref{eq:LPMsupp}).

The $\epsilon \sim (K \cdot \Delta X)^2$ suppression factor
favors larger photon angles relative to $\Delta\x$ over smaller ones.
So it's important to review what sets the upper limit on photon angle
in the current context.
The difference (\ref{eq:collinear2}) between the bremsstrahlung
probabilities in medium and in vacuum will sometimes be positive
and sometimes negative.  Consider the average of
(\ref{eq:collinear2}) over rotations of the direction of
$\v_{\rm f}$ around the axis defined by $\v_1$.
Let $\theta_{\gamma 1}$ and $\theta_{{\rm f}1}$ be the angles that
$\k$ and $\v_{\rm f}$ makes with $\v_{\rm 1}$, and let
$\phi_{{\rm f}1}$ represent the azimuthal angle being averaged
over.
Ignore mass effects, so that $\v_{\rm f}$ is a unit vector.
The averaging
gives
\begin {equation}
  \left\langle \frac{\varepsilon\cdot V_{\rm f}}{K \cdot V_{\rm f}}
  \right\rangle_{\rm\phi_{{\rm f}1}}
  =
  \frac{\varepsilon\cdot V_1}{K \cdot V_1}
  \qquad \mbox{if $\theta_{\gamma 1} > \theta_{{\rm f}1}$} ,
\end {equation}
where $\langle\cdots\rangle_{\phi_{{\rm f}1}}$ indicates
averaging over $\phi_{{\rm f}1}$.
So the corresponding average of the medium effect (\ref{eq:collinear2})
vanishes if $\theta_{\gamma 1}$ is larger than the deflection angle
$\Delta\theta = \theta_{{\rm f}1}$ of the charged particle trajectory
in Fig.\ \ref{fig:collinear}.  For multiple scatterings involving
a number of consecutive trajectory directions $\v_1$, $\v_2$, ...,
$\v_N = \v_{\rm f}$, one can start by averaging over rotations of
$\v_N$ about $\v_{\rm N-1}$ and then work back recursively to
show that a similar cancellation occurs if
$\theta_{\gamma,n{-}1} > \theta_{n,n{-}1}$ for all $n$.

\begin {figure}
\begin {center}
  \includegraphics[scale=0.3]{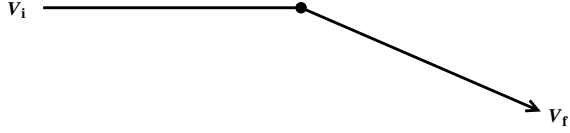}
  \caption{
    \label {fig:isolated}
    An isolated collision.
    }
\end {center}
\end {figure}

Eqs.\ (\ref{eq:collinear}) and
(\ref{eq:collinear2}) also show the behavior associated with collinear
logs discussed in Section \ref{sec:collinear}.
Compare (\ref{eq:collinear2}) to the usual calculation of
bremsstrahlung from an isolated collision like Fig.\ \ref{fig:isolated}:
\begin {equation}
  |\varepsilon \cdot J|^2
  \propto
  \left|
    - \frac{\varepsilon\cdot V_{\rm i}}{K \cdot V_{\rm i}}
     + \frac{\varepsilon\cdot V_{\rm f}}{K \cdot V_{\rm f}}
  \right|^2 .
\label {eq:isolated}
\end {equation}
The collinear divergences associated with photons collinear with
the initial or final direction correspond to the divergence of
(\ref{eq:isolated}) as $K \cdot V_{\rm i}$ or $K \cdot V_{\rm f}$
vanishes, respectively.  If polarizations are summed over, this
corresponds to divergences like $1/\theta^2_{{\rm i}\gamma}$ and
$1/\theta^2_{{\rm f}\gamma}$, where $\theta_{{\rm i}\gamma}$
and $\theta_{{\rm f}\gamma}$ are the corresponding angles.
Integration over angles $d^2\Omega$ then gives the usual collinear
logarithmic divergences.  In (\ref{eq:collinear}), however,
the cancellation with the vacuum term softens the
$K \cdot V_{\rm f} \to 0$ behavior, so that there is no
corresponding logarithmic divergence in the angular integration.
The other divergence, $K\cdot V_{\rm i}\to 0$, is cut off
once
$K\cdot\Delta X \propto K\cdot V_{\rm i}$ becomes small enough
that $1-\cos(K\cdot\Delta X) \sim (K\cdot\Delta X)^2$
in (\ref{eq:collinear2}): that is, when
$K \cdot \Delta X \ll 1$.

\subsection {The case \boldmath$Q_\perp \ll (Q_\perp)_{\rm typ}$}
\label {app:smallQT}

The diagram of Fig.\ \ref{fig:smallQT} corresponds to
\begin {align}
   J^\mu(K) &\propto
   - \frac{V_{\rm i}^\mu}{K \cdot V_{\rm i}} \, e^{i K\cdot X_1}
   + \frac{V_1^\mu}{K \cdot V_1}
              (e^{i K\cdot X_1}-e^{i K\cdot X_2})
   + \frac{V_{\rm f}^\mu}{K \cdot V_{\rm f}} \, e^{i K \cdot X_2}
\nonumber\\
  &= e^{iK\cdot X_2}
  \left( \frac{V_1^\mu}{K \cdot V_1}
          - \frac{V_{\rm i}^\mu}{K \cdot V_{\rm i}} \right)
  (e^{-i K \cdot \Delta X} - 1) ,
\end {align}
where $V_1$ is the intermediate particle direction and the
initial and final directions are equal: $V_{\rm i}=V_{\rm f}$.
So
\begin {equation}
  |\varepsilon \cdot J|^2
  \propto
  \left|
      - \frac{\varepsilon\cdot V_{\rm i}}{K \cdot V_{\rm i}}
     + \frac{\varepsilon\cdot V_1}{K \cdot V_1}
  \right|^2
  \left| e^{-i K \cdot \Delta X} - 1 \right|^2 .
\label {eq:smallQT2}
\end {equation}
The factor $|e^{-iK\cdot\Delta X}-1|^2$ gives
$\epsilon\sim(K\cdot\Delta X)$ LPM suppression when
$K\cdot\Delta X \ll 1$.
The other factor in (\ref{eq:smallQT2}) looks just like
the result (\ref{eq:isolated}) that the individual collisions would each
give if they were isolated.
The two terms in this factor approximately cancel each other
only when the photon angle is large compared to the angle
between $V_{\rm i}=V_{\rm f}$ and $V_1$.
In order of magnitude, the characteristic angle $\theta$ of
bremsstrahlung from two collisions with
canceling deflections is therefore similar to that of two
collisions with deflections in the same direction.


\subsection {Collinear logarithms for thick media}
\label {app:thicklog}

Here I will flesh out the argument for the collinear logarithm
(\ref{eq:thicklog}) that appears for the high-$Q_\perp$ tail in
Fig.\ \ref{fig:qedPbrem}b in the thick medium case
$L \gg \linf(\omega)$.  Return to Eq.\ (\ref{eq:thetai2}).
Multiple scatterings preceding the rare, larger-than-typical angle
scattering will only affect the bremsstrahlung process
if they occur within the corresponding mean free time
$\lf(\theta_{{\rm i}\gamma}) \sim 1/\omega\theta_{{\rm i}\gamma}^2$.
So, the condition (\ref{eq:thicklog}) should more accurately be
written
\begin {equation}
   \max\left(
      \sqrt{ \frac{1}{\omega L} } \,,
      \sqrt{ \frac{\hat q \min[L,\lf(\theta_{{\rm i}\gamma})] }{E^2} } 
   \right)
   \lesssim \theta_{{\rm i}\gamma}
   \lesssim \Delta\theta .
\label {eq:thetai3}
\end {equation}
For $L \gg \linf(\omega)$, the lower bound on $\theta_{{\rm i}\gamma}$
which dominates is
\begin {equation}
   \sqrt{ \frac{\hat q \, \lf(\theta_{{\rm i}\gamma}) }{E^2} } 
   \sim
   \sqrt{ \frac{\hat q}{\omega E^2 \theta_{{\rm i}\gamma}^2} } 
   \lesssim \theta_{{\rm i}\gamma} .
\end {equation}
This is equivalent to
\begin {equation}
   \left( \frac{\hat q}{\omega E^2} \right)^{1/4}
   \lesssim \theta_{{\rm i}\gamma} .
\end {equation}
Using (\ref{eq:linfQED}), the constraint (\ref{eq:thetai3})
can in this case be written in the form
\begin {equation}
   \frac{[\hat q \, \linf(\omega)]^{1/2}}{E}
   \lesssim \theta_{{\rm i}\gamma}
   \lesssim \Delta\theta
   \sim \frac{Q_\perp}{E}
   \qquad
   \mbox{for $L \gg \linf(\omega)$.}
\end {equation}
A significant range exists when $Q_\perp^2 \gg \hat q \, \linf(\omega)$, and
the corresponding logarithm is (\ref{eq:thicklog}).


\section {Collinear logarithms in QCD}
\label{app:qcd}
\label{app:QCDcollinear}

\begin {figure}
\begin {center}
  \includegraphics[scale=0.3]{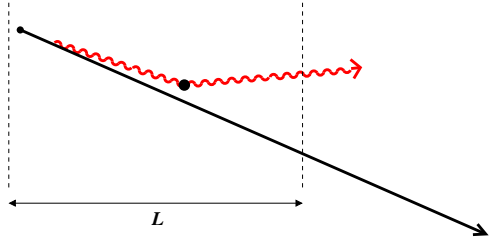}
  \caption{
     \label{fig:collinearQCD}
     A QCD addition to the diagrams of Fig.\ \ref{fig:collinear}a--b.
  }
\end {center}
\end {figure}

In the case of QCD, the gluon scatters from the medium.  If we
again focus on the case of a single significant scattering, then
Fig.\ \ref{fig:collinearQCD} needs to be added to the situation
considered for QED in Figs.\ \ref{fig:collinear}a--b.
For simplicity, neglect the original creation of the particle at the left-hand
side of these diagrams and instead considers it to come from
infinity.  This is the gluon bremsstrahlung situation analyzed
long ago by Gunion and Bertsch \cite{GunionBertsch}.
Schematically, the amplitude for bremsstrahlung compared to the
amplitude for scattering without bremsstrahlung is
(in the high energy limit)
proportional to \cite{GunionBertsch}%
\footnote{
   My $\q_\perp$ and $\k_\perp$ are respectively the
   ${\bm l}_\perp$ and $\q_\perp$ of Ref.\ \cite{GunionBertsch}.
}
\begin {equation}
  g T^a T^b \,
  \frac{(\k_\perp - x \q_\perp)\cdot {\bm\varepsilon}_\perp (1-x)}
       {| \k_\perp  - x \q_\perp|^2}
  - g T^b T^a
  \frac{\k_\perp\cdot {\bm\varepsilon}_\perp (1-x)}
       {\k_\perp^2}
  - g [T^a,T^b] \,
  \frac{(\k_\perp - \q_\perp)\cdot {\bm\varepsilon}_\perp (1-x)}
       {| \k_\perp - \q_\perp |^2} ,
\label {eq:GB}
\end {equation}
where $\perp$ is defined relative to the initial particle direction,
$\q_\perp$ and $a$ characterize the transverse momentum transfer
and adjoint color index associated with the collision, and $\k_\perp$
and $b$ characterize the transverse momentum and adjoint color index
of the final bremsstrahlung gluon.  In the corresponding QED
calculation, the first two terms cancel in the limit that the
photon angle is large compared to the deflection angle of the
charged particle ($k_\perp \gg x q_\perp$).
In QCD, they do not because the color generators
do not commute.  Instead, in the soft gluon case, the cancellation
occurs
between all three terms of (\ref{eq:GB}), when the bremsstrahlung
gluon angle becomes large compared to the deflection
of the gluon due to $\q_\perp$ ($k_\perp \gg q_\perp$).
That is, in the notation of (\ref{eq:dtheta2}), bremsstrahlung
gluons associated with the collision decouple when
$\theta \gg (\Delta\theta)_{\rm g}$.

The amplitude proportional to (\ref{eq:GB}) diverges when any of
the denominators goes to zero, leading to collinear divergences.
Now return from the Gunion and Bertsch problem back to the
case of Figs.\ \ref{fig:collinear} and \ref{fig:collinearQCD}
where the original particle is
created at the beginning of the trajectories shown.
The collinear divergence corresponding to Fig.\
\ref{fig:collinear}a, i.e.\ the square of the first term
in (\ref{eq:GB}), will cancel against the corresponding
virtual correction to bremsstrahlung in the vacuum case.
The rest will give
a collinear logarithm that will be cut off at small angles
when the LPM formation
time becomes of order $L$, just as in the QED case.
One way to understand this is to note that collinear divergences
correspond to the intermediate particle states in Figs.\
\ref{fig:collinear}a--b and \ref{fig:collinearQCD} going on shell.
However, for the case of Figs.\ \ref{fig:collinear}b and
\ref{fig:collinearQCD}, the intermediate particle lines can
have length at most $L$, which introduces an uncertainty in their
energy of order $1/L$.  This provides a lower bound to how on-shell
those particles can be and so cuts off the corresponding collinear
divergences.


\section {Contribution to \boldmath$\Delta E$ from \boldmath$x\to1$}
\label{app:xto1}

The HO formula (\ref{eq:dEHO}) for $\Delta E$ comes from taking the
small $x$ approximation to the spectrum (\ref{eq:IHO}):
\begin {equation}
   \omega \, \frac{d}{d\omega}(I-I_{\rm vac})_{\rm HO} \to
   \frac{2 C_s \alpha}{\pi} \,
   \ln\left| \cos\sqrt{\frac{-i \qhatA L^2}{2\omega}} \right| .
\label {eq:x1}
\end {equation}
In the thin-media limit, integration over $\omega$ is dominated by
small $x$ of order
\begin {equation}
   x \sim \frac{\qhatA L^2}{E} \sim \frac{L^2}{\Linf^2} .
\end {equation}
Integrating (\ref{eq:x1}) over $\omega$ from zero to infinity
gives (\ref{eq:dEHO}).
But this procedure implicitly ignores the possibility of an additional
contribution from small $1{-}x$ in the case of bremsstrahlung
from a quark (${\rm q}{\to}{\rm gq}$) or anti-quark.

Define
\begin {equation}
   \Delta E_{\rm q}
   \equiv 
   \int_0^E d\omega \> \omega \, \frac{d}{d\omega}(I-I_{\rm vac})
\end {equation}
to be the average energy loss of the quark in ${\rm q}{\to}{\rm gq}$.
In the thin media limit, one contribution to $\Delta E_{\rm q}$ is
the small $x$ approximation just discussed.  But there is another
contribution from small $1{-}x$.  In the limit of small $1{-}x$,
the spectrum (\ref{eq:IHO}) becomes
\begin {equation}
   \omega \, \frac{d}{d\omega}(I-I_{\rm vac})_{\rm HO} \to
   \frac{\cf \alpha}{\pi} \,
   \ln\left| \cos\sqrt{\frac{-i \qhatF L^2}{2(1-x)E}} \right| .
\label {eq:x2}
\end {equation}
The $\omega$ integral of (\ref{eq:x2}) is dominated by small $1{-}x$
of order
\begin {equation}
   1-x \sim \frac{\qhatF L^2}{E} ,
\label {eq:x3}
\end {equation}
and the integration gives an small $1{-}x$ contribution of
$\tfrac18 \cf \alpha \qhatF L^2$ to $\Delta E_{\rm q}$.
Adding this to the small-$x$ contribution of (\ref{eq:dEHO}),
\begin {equation}
   (\Delta E)_{\rm q,HO} \simeq
   \tfrac14 \cf \alpha (\qhatA+\half\qhatF) L^2 .
\end {equation}

Now instead consider the average energy loss of the leading parton for
${\rm q}{\to}{\rm gq}$, which is
\begin {equation}
   \Delta E_{\rm lead}
   \equiv 
   \int_0^{E/2} d\omega \> \omega \, \frac{d}{d\omega}(I-I_{\rm vac})
   +
   \int_{E/2}^E d\omega \> (E-\omega) \, \frac{d}{d\omega}(I-I_{\rm vac}) .
\end {equation}
The replacement of the $\omega$ factor by $E{-}\omega$ in the second term
produces an additional suppression of (\ref{eq:x3}) to the
small $1{-}x$ contribution.  As a result, the small $x$ contribution
dominates in the thin media limit, and so $\Delta E_{\rm lead}$ is
simply the $\Delta E$ quoted in the main text.

If one is interested in understanding the parametric dependence of
the ${\rm q}{\to}{\rm gq}$ bremsstrahlung spectrum for the case of
small $1{-}x$, it is easy to adapt the QCD results of the main text.
In this case, the final-state quark is the particle that is most
easily scattered and so the one whose scattering sets the scale of
the LPM effect.  Parametrically, the result for
the bremsstrahlung probability when $\omega > E/2$
will look just like
Fig.\ \ref{fig:qcdPbremL} but with the replacement
\begin {equation}
   \linf(\omega) \to \sqrt{\frac{E-\omega}{\qhatF}}
\end {equation}
instead of (\ref{eq:linfQCD}).


\end{document}